\documentclass{aa}
\usepackage[utf8]{inputenc}
\usepackage[varg]{txfonts}
\bibpunct{(}{)}{;}{a}{}{,}

\newcommand{\disperse}{DisPerSE$\:$}

\title{Like a spider in its web: a study of the Large Scale Structure around the Coma cluster}
\author{Nicola Malavasi\inst{1}
\and
Nabila Aghanim\inst{1}
\and
Hideki Tanimura\inst{1}
\and
Victor Bonjean\inst{1}
\and
Marian Douspis\inst{1}}

\institute{Institut d'Astrophysique Spatiale, CNRS, Universit{\'e} Paris-Sud, B{\^a}timent 121, Orsay, France}

\date{Received: 1st January 2019 / Accepted: 1st January 2019}

\abstract{The Cosmic Web is a complex network of filaments, walls and voids that represent the largest structures in the Universe. In this network, which is the direct result of structure formation, galaxy clusters occupy central positions as the nodes, connected through the filaments.
In this work, we investigate the position in the Cosmic Web of one of the most known and best studied clusters of galaxies, the Coma cluster.
We make use of the Sloan Digital Sky Survey Data Release 7 (SDSS DR7) Main Galaxy Sample and of the Discrete Persistent Structure Extractor (\disperse) to detect large scale filaments around the Coma cluster and we analyse the properties of the Cosmic Web. We study the network of filaments around Coma in a 75 Mpc radius region. We find that the Coma cluster has a median connectivity of 2.5, in agreement with measurements from clusters of similar mass in the literature, as well as with what expected from numerical simulations. It is indeed connected to 3 secure filaments which connect Coma to Abell 1367 and to several other clusters in the field. The location of these filaments in the vicinity of Coma is consistent with features detected in the X-ray, as well as the likely direction of infall of galaxies, as for example NGC4839. The overall picture that emerges of the Coma cluster is that of a highly connected structure occupying a central position as a dense node of the Cosmic Web. We also find a tentative detection, at 2.1$\sigma$ significance, of the filaments in the SZ signal.}

\titlerunning{Like a spider in its web}
\authorrunning{Malavasi et al.}


\begin{document}

\maketitle

\section{Introduction}
The spatial distribution of galaxies in the Universe is far from being uniform at the megaparsec scale. Ever since the advent of wide area spectroscopic surveys such as the Two-Degree Field Galaxy Redshift Survey \citep[2dFGRS,][]{Colless2001}, the Sloan Digital Sky Survey \citep[SDSS,][]{York2000}, the Galaxy And Mass Assembly survey \citep[GAMA,][]{Driver2009}, or the Vimos Public Extragalactic Redshift Survey \citep[VIPERS,][]{Scodeggio2018}, multiple demonstrations have been provided that galaxies form a complex network of structures, often referred to as the Cosmic Web.

The Cosmic Web \citep{deLapparent1986, Bond1996} is an intricate system of features, composed of elongated, hundreds of Mpc long one-dimensional structures (called filaments) and flat, planar, two-dimensional walls. Walls surround vast, empty regions called voids. In this complex picture, clusters (often also referred to as nodes) occupy peculiar places at the intersection of filaments \citep[e.g.][]{AragonCalvo2010, Cautun2014, Malavasi2017}, although smaller, lower-mass galaxy associations can be found inside the filaments, referred to as knots.

This configuration of matter arises from the process of structure formation driven by gravitational collapse: following the initial density anisotropies whose imprint is visible in the Cosmic Microwave Background \citep[CMB, e.g.][]{PlanckCollaborationMap} matter departs from under-dense regions which become voids, flows inside walls to the filaments which are located at the intersections, and finally flows inside filaments to reach clusters. It is in this evolving environment that galaxies are formed and that complex interactions with the gaseous phase present in these structures take place.

Detecting the Cosmic Web from the galaxy distribution is not an easy task and many methods to achieve it have been proposed \citep[see e.g.][for a comparison of some of the methods]{Libeskind2018}. These methods often rely on the study of the geometry of the galaxy density field or of the tidal field to obtain information on its topology. Despite difficulties, features of the Cosmic Web have been successfully identified in both simulations \citep[see e.g.][]{AragonCalvo2010, Cautun2014, Chen2015Simu, Laigle2018, Kraljic2019Simu} and galaxy surveys \citep[see e.g.][]{Tempel2014, Chen2016SDSS, Malavasi2017, Laigle2018, Kraljic2018GAMA}. It is now starting to become possible to study the Cosmic Web in great detail.

Nevertheless, while features of the Cosmic Web such as filaments have been identified starting from galaxy surveys, the detection of their gaseous content has proved to be much more challenging. Indeed, detection of the Warm-Hot Intergalactic Medium (WHIM) in filaments has been obtained through direct observations of the X-ray emission or of the Sunyaev-Zel'dovich effect \citep[SZ,][]{SunyaevZeldovich1969} signal in bridges of matter between galaxy clusters \citep[e.g.][]{Akamatsu2017, Bonjean2018} as well as with stacking procedures \citep[see e.g.][]{Tanimura2019}.

On the other hand, cluster physics is now starting to be well understood. Clusters of galaxies have been thoroughly investigated in terms of their galaxy and gas distribution. For some of them, a detection of filamentary structure departing from the cluster itself and outlining the position that clusters occupy as nodes at the intersection of filaments of the Cosmic Web has been achieved \citep[see e.g.][]{Sanders2013, Eckert2015}.

In this regard, a very important observable that can be measured for clusters is the so-called connectivity \citep[see e.g.][and references therein]{Colombi2000,Codis2018}. It is a measurement of the number of filaments connected to galaxy clusters and it is known to scale with cluster/group mass \citep{Sarron2019, DarraghFord2019}. Given that events, such as cluster mergers, may influence the number of connections to a cluster, the connectivity is an important test for structure evolution.

Among the most studied clusters, the Coma cluster has a central place. Known since the early works of \citet{HubbleHumason1931}, it is one of the best known and most thoroughly investigated structures in the nearby Universe \citep[for a historical review on the modern investigation of the Coma cluster of galaxies see][]{Biviano1998}. It is located in the nearby Universe ($z = 0.023$ \citealt{PlanckCollaborationXComa}) at the centre of the northern galactic region, which makes it a suitable target for a large number of optical observations. Indeed, in the years it has received coverage at multiple wavelengths such as for example X-ray \citep[both through the ROSAT All-Sky survey and with the XMM-Newton satellite, see][]{Briel1992,Neumann2001,Neumann2003}, SZ signal \citep{PlanckCollaborationXComa}, radio \citep{BrownRudnick2011}, as well as optical spectroscopy to precisely identify member galaxies \citep[see e.g.][]{Adami2005,denBrok2011}.

The Coma cluster is known to have a complex morphology, including a high degree of substructure \citep{Biviano1996, Adami2005, Adami2009PFA} and an elongated shape in the gas phase, visible in the SZ signal \citep{PlanckCollaborationXComa} and in the X-ray, where filamentary structures in the core of the cluster are also visible \citep{Sanders2013}. In fact the Coma cluster is likely connected to other structures in the field, including several Abell clusters, the closest one being A1367 \citep{West1998,Mahajan2018}. Coma also presents an infalling group of galaxies, close to the position of NGC 4839, about half a degree South-West of the main cluster core \citep[see e.g.][and references therein]{Biviano1998, Neumann2001, PlanckCollaborationXComa}.

Although the general picture is known and the connections of Coma to other clusters have been highlighted, still a complete, quantitative picture of the position of Coma in the context of the Cosmic Web is missing. The goal of this study is to explore the place that the Coma cluster occupies in the Large Scale Structure surrounding it (on a scale of $\sim 100\: \mathrm{Mpc}$) and in the Cosmic Web. We quantitatively investigate the connections of the Coma cluster to other structures, the filaments connected to the cluster and its connectivity.

The paper is organised as follows: in Section \ref{data}, we describe the data sets that we have used to detect the Cosmic Web around Coma. In Section \ref{method}, we briefly describe the DisPerSE algorithm that we have applied to the SDSS survey. In Section \ref{results_connections}, we describe our results in terms of the connectivity of the Coma cluster and its connections to other clusters in the field, while in Section \ref{results_gasphase}, we present a tentative detection of the filaments in the gas phase through the use of the SZ signal. We discuss our results in Section \ref{discussion} and summarise our conclusions in Section \ref{conclusions}. Throughout this paper we use a \citet{PlanckCollaborationXIIICosmoparams} cosmology, with $H_{0} = 67.74\: \mathrm{km}\: \mathrm{s}^{-1}\: \mathrm{Mpc}^{-1}$, $\Omega_{m} = 0.3075$, $\Omega_{\Lambda} = 0.6925$. Equatorial coordinates are given in the J2000 reference.

\section{Data}
\label{data}
The data set that we used for this work is composed of a large-area galaxy survey (to detect filaments) and of several samples of galaxy clusters and groups (detected at various wavelengths, from the X-ray to optical and microwaves), which we used to study the connections of the Coma supercluster system on a large scale. 

\subsection{Optical galaxies}
\label{data_galaxies}
The catalogue of optical galaxies which we used to detect the filaments is the 7th Data Release (DR7) of the Sloan Digital Sky Survey \citep[SDSS][]{York2000, Abazajian2009}. In particular, we made use of the Main Galaxy Sample\footnote{available at: \url{http://classic.sdss.org/legacy/index.html}} (MGS) of the Legacy Survey \citep{Strauss2002}, a uniformly selected sample of galaxies in the redshift range $z \sim 0 \div 0.3$ limited at $r$-band Petrosian magnitude $r_{\mathcal{P}} \leq 17.77$. We chose this sample as it presents a rather well understood selection function, a high degree of completeness as well as a large and uniform area coverage in the redshift range of interest for the Coma supercluster system. The number of galaxies in the MGS is $697\,920$.

Due to the Coma supercluster position and the fact that the Southern region only consists of three stripes where spectroscopic information is available, we restrict our analysis to the contiguous area of the north region. Following \citet[][]{Strauss2002}, we select only galaxies with a secure redshift measurement (\textsc{zwarning} = 0, \textsc{zconffinal} > 0.35, \textsc{zfinal} > 0). The final number of sources available for filament detection is $566\,452$.

The volume density of sources as a function of redshift in the survey area (hereafter referred to as $n(z)$) presents a peak in the redshift range $z = 0.01 \div 0.04$ followed by a steady but shallow decline. Therefore the variation with redshift of the mean inter-galaxy separation ($\langle D_{z} \rangle = 1/\sqrt[3]{n(z)}$) is slow and the properties of the filaments such as their length will not vary drastically across the survey volume. Moreover, the peak in the $n(z)$ at $z = 0.01 \div 0.04$ (i.e. in the redshift range covered by our $\sim 150$ Mpc across region of interest) is entirely due to the Coma supercluster system. If the member galaxies are removed, the volume density of sources becomes flat in the considered redshift range, resulting in no variation in the density of tracer galaxies in the region of interest.

The angular distribution of galaxies on the plane of the sky, shows a rather uniform and homogeneous coverage, without gaps or holes and without strong unevenness across the survey area. The only detectable anisotropy is a small dependance of the surface density of galaxies with declination. After the extraction of the skeleton (described in Section \ref{method}), we visually compare the filaments in three declination slices, finding no major difference. Also the length distributions for the filaments in the three declination slices show no alteration.

We used the \disperse algorithm to detect filaments on all galaxies in the redshift range $0 \lesssim z \lesssim 0.3$. We refer to Sect. \ref{method} and to Malavasi et al. (in preparation) for a more detailed description of the procedure.

\subsection{Cluster catalogues}
\label{data_cluster}
In order to confirm our filament detection and to study the LSS around the Coma supercluster system, we made use of several catalogues of groups and clusters of galaxies. These catalogues have been constructed from observations at various wavelengths, namely in the X-ray, optical, and in the microwave domain.

\subsubsection{The MCXC cluster catalogue}
The Meta-Catalogue of X-ray detected Clusters of galaxies \citep[MCXC,][]{Piffaretti2011} is a homogenised compilation of a large number of pre-existing galaxy cluster catalogues based on serendipitous discoveries and on the ROSAT All-Sky Survey \citep{Voges1999}. The MCXC catalogue provides measurements of $L_{500}$ in the band $0.1-2.4$ keV, $M_{500}$ and $R_{500}$ for 1743 clusters in the redshift range $z \sim 0 \div 1$ and on the full sky extension.

\subsubsection{The \emph{Planck} cluster catalogue}
We make use of the publicly available SZ cluster database\footnote{\url{http://szcluster-db.ias.u-psud.fr/}}. This catalogue is the union of several catalogues of clusters detected via their SZ signal with several instruments: \emph{Planck} \citep[the early, first, and second \emph{Planck} catalogue of SZ sources, ESZ, PSZ1, PSZ2,][]{PlanckCollaborationVIIIESZ, PlanckCollaborationXXIXClusters, PlanckCollaborationXXVIIClusters}, the South Pole Telescope \citep[SPT,][]{Williamson2011, Reichardt2013, Ruel2014, Bleem2015}, the Atacama Cosmology Telescope \citep[ACT,][]{Hasselfield2013}, the Arcminute Microkelvin Imager \citep[AMI,][]{AMIConsortium2012, AMIConsortium2013, AMIConsortium2013bis, Schammel2013}, and the Combined Array for Research in Millimeter-wave Astronomy \citep[CARMA,][]{Brodwin2015, Buddendiek2015}. The catalogue comprises 2676 sources, of which 1748 are clusters with a confirmed redshift distributed over the full extent of the sky. Masses are determined through the use of the $Y_{500}-M_{500}$ relation by \citet{PlanckCollaborationXXYM}.

\subsubsection{The SDSS optical groups catalogue}
Analysing the SDSS Data Release 12 \citep[SDSS DR12,][]{Eisenstein2011, Alam2015}, \citet{Tempel2017} ran a Friend-of-Friend algorithm on the galaxy distribution and identified groups and clusters in the redshift range $z \sim 0 \div 0.2$. Of the $88\,662$ groups and clusters they identify, many consists of galaxy pairs or have a very small mass ($68\,887$ have a number of galaxies belonging to the group $N_{gal} \leq 3$). We eliminate from the sample all clusters with a number of galaxies belonging to the group below $N_{gal} = 6$, resulting in 6873 objects. The richness threshold was chosen following \citet{Tempel2017} in order to select only clusters with a reliable mass estimate. The resulting mass range is $10^{11} \div 10^{15} M_{\sun}$.

\subsubsection{The SDSS supercluster catalogue}
We use the supercluster catalogue obtained in the SDSS DR7 \citep{Abazajian2009} by \citet{Liivamaagi2012} who identified superclusters in the galaxy distribution by means of both an adaptive local threshold and a fixed global threshold in the luminosity density field. These superclusters are defined as extended objects with a complex shape. We consider here the superclusters identified in the fixed threshold case (with the value of the threshold set to a normalised density of 5.0, see \citealt{Liivamaagi2012} for more detail). There are 982 distinct superclusters in the sample.

\subsection{The Coma Cluster}
\label{data_coma}
As already stated in the introduction, the Coma cluster has been the subject of several investigations, including substructure and its connection to other clusters in the field. In our analysis, we assume a virial radius size for Coma of $\mathrm{R}_{vir} = 97\arcmin$ \citep{LokasMamon2003}, which at the redshift of Coma ($z = 0.023$) corresponds to $\mathrm{R}_{vir} = 2.86$ Mpc. When comparing the connectivity values with other measurements from the literature, we use a value of $M_{200}$ for Coma of $M_{200} = 5.3 \cdot 10^{14} M_{\sun}$ \citep{GavazziComa2009} and a value for the total virial mass (including dark matter) of $M_{vir} = 1.4 \cdot 10^{15} M_{\sun}$ \citep{LokasMamon2003}. For the centre of the Coma cluster we use the coordinates of NGC4874, an elliptical galaxy considered to lie at the centre of the cluster \citep[see][]{LokasMamon2003} and located at $(\alpha_{J2000}, \delta_{J2000}) = (194.899\deg, 27.959\deg)$. We use these coordinates for all our analysis except when investigating the filaments in the SZ, where the coordinates of the SZ centroid, provided by \citet{PlanckCollaborationXComa} and located at $(\alpha_{J2000}, \delta_{J2000}) = (194.946\deg, 27.931\deg)$, are preferred (e.g. for subtracting the profile of Coma from the SZ map, see Section \ref{results_gasphase}). The information on these quantities is summarised in Table \ref{coma_values}.

\begin{table*}
\caption{Physical quantities for the Coma cluster used in this paper.}
\label{coma_values}
\centering
\begin{tabular}{c c c}
\hline\hline
Quantity & Value & Reference \\
\hline
Redshift, z & 0.023 & \citet{PlanckCollaborationXComa} \\ 
Virial Radius (arcmin), $\mathrm{R}_{vir}$ & 97 & \citet{LokasMamon2003} \\
$M_{vir} ( M_{\sun})$ & $1.4 \cdot 10^{15}$ & \citet{LokasMamon2003} \\
$M_{200} (M_{\sun})$ & $5.3 \cdot 10^{14} $ & \citet{GavazziComa2009} \\
Centre (galaxies, deg) (R.A., Dec.) & (194.899, 27.959) & \citet{LokasMamon2003} \\ 
Centre (SZ emission, deg) (R.A., Dec.) & (194.946, 27.931) & \citet{PlanckCollaborationXComa} \\ 
\hline
\end{tabular}
\end{table*}

\subsection{The \emph{Planck} $y$-map}
\label{planck_map_data}
In order to investigate the detection of the filaments around the Coma cluster in the gas phase through the SZ effect, we need a map of the Compton $y$ signal on the plane of the sky in the Coma region.

We use the full-sky $y$-map from the 2015 \emph{Planck} data release\footnote{\url{https://pla.esac.esa.int}}. The full-sky $y$-map is available in HEALPix\footnote{\url{https://healpix.sourceforge.io}} format ($N_{side} = 2048$, \citealt{Gorski2005}). Following \citet{Tanimura2019Superclusters}, of the various maps available, constructed with different methods (e.g. NILC, \citealt{Remazeilles2013} or L-GMCA, \citealt{Bobin2016}), we make use of the $y$-map constructed from the combination of the multi-band frequency maps by 
\citet{PlanckCollaborationXXIISZMap} with the MILCA method \citep{Hurier2013}. 

To mask for the emission of contaminating sources in our analysis, we rely on the masks that accompany the $y$-maps in the 2015 \emph{Planck} data release. We mask 40\% of the sky around the galactic plane as well as IR and radio point sources \citep{PlanckCollaboration2016XXVIPCCS2}. Moreover, in order not to be biased in our analysis by the SZ signal coming from galaxy clusters, we mask clusters and groups identified in the SZ, X-ray and optical, as listed in \citet{Tanimura2019Superclusters}. These include the already mentioned \emph{Planck} SZ clusters and the MCXC X-ray clusters. To these samples are added $26\,111$ clusters from the redMaPPer sample \citep{Rykoff2014}, $158\,103$ clusters from the WHL sample \citep{Wen2012, Wen2015}, and $46\,479$ clusters from the AMF sample \citep{Banerjee2018}. All clusters are masked out to a distance of $3 \times R_{500}$ from the cluster centre on the plane of the sky, with the only exception of the Coma cluster. In order to be able to perform a more accurate analysis in the close proximity of Coma, the cluster itself is not masked, neither is any other cluster inside a cylinder of radius $3 \times R_{500}$ from the centre of the SZ emission of Coma on the plane of the sky (see Table \ref{coma_values}) and height $\pm 75 \mathrm{Mpc}$ centred at the redshift of Coma on the Line of Sight direction.

\section{Method}
\label{method}
We detected filaments in the MGS by making use of the Discrete Persistent Structure Extractor \citep[\disperse\footnote{\url{http://www2.iap.fr/users/sousbie/web/html/indexd41d.html}}][]{Sousbie2011a,Sousbie2011b}. As a complete description of the filament detection procedure and of the Large-Scale Structure catalogue detected in the SDSS DR7 (and DR12) will be provided in a dedicated paper (Malavasi et al., in preparation), here we briefly summarise the code characteristics and a few details of its application to the data.

The \disperse code detects filaments as ridges of the density field (i.e. field lines running along paths of constant gradient). It can be applied to a density field computed starting from the distribution of galaxy positions in the 3D space.

We computed the density field of the galaxies in the MGS using the Delaunay Tessellation Field Estimator \citep[DTFE,][]{SchaapWeygaert2000,WeygaertSchaap2009}. The DTFE provides a way to smooth the measurement of the density by averaging the value of the density at the position of each galaxy with the values of the density of the galaxies directly connected to it through tetrahedrons of the tessellation. This process can be iterated by averaging again the values of the density which have been already averaged once. In our case, we chose three levels of smoothing: no smoothing of the density field, one iteration of smoothing and two iterations of smoothing (hereafter 1-smooth and 2-smooth, respectively). We expect the no-smoothing case to be more sensitive to the shot noise and to present a larger number of small scale variations of the density. In this case, the number of cosmic web features (such as filaments) will be higher, with a corresponding larger fraction of possible local or spurious features. On the other hand in the other extreme case of 2-smoothing, the density field will loose a large number of small scale variations and detecting maxima and minima of the density field, together with filaments, will become more difficult. In the rest of this paper, we often refer to the intermediate 1-smoothing case as the fiducial one, and we discuss what would happen if the two other extreme cases were considered.

When \disperse is applied, it uses the discrete Morse theory to find critical points of the density field (maxima, saddles and minima). These points are then connected by means of field lines and paired in topological constructs called persistence pairs by means of the persistent homology theory. Each persistence pair is ranked based on the relative density contrast of the critical points composing it (i.e. its persistence value). The distribution of persistence values for the pairs of critical points detected in the data is compared to the same distribution for pairs of critical points detected in a Gaussian random field, to determine which ones are due to sampling noise.

A threshold can then be set in the comparison by eliminating, through the process of topological simplification, all the persistence pairs which are closer than a certain number of $\sigma$ to the noise distribution. In this analysis, we considered different persistence thresholds, namely a 3$\sigma$ and a 5$\sigma$ persistence threshold. While the choice of the persistence threshold is somewhat arbitrary, a lower threshold ensures that all the structures with a level of significance smaller than $3\sigma$ with respect to noise are eliminated from the sample. A $5\sigma$ persistence threshold is a more conservative cut, which largely sacrifices statistics to keep only the most secure features of the Cosmic Web. In the following, we will favour the $3\sigma$ persistence threshold to guarantee a good balance between number of detected features and significance, while checking that the trends we recover are maintained if we use the more conservative $5\sigma$ persistence cut.

As a final remark, we also applied a smoothing to the position of the skeleton segments, by averaging the positions of a segment's extrema with those of the extrema of contiguous segments. This additional smoothing is performed to mitigate the effect of shot noise on the shape of the filaments, which would otherwise induce very sharp changes of direction and would imply unphysical edges at the position where the segments are joined. However, as the sampling of SDSS galaxies is high enough and the segments composing the filaments are rather short, the impact of such a smoothing on our results is minimal.

\section{The Coma cluster in the Cosmic Web}
\label{results_connections}
This work presents an analysis of the LSS around the Coma cluster. The LSS is defined both in terms of filaments directly attached to the Coma cluster, as well as the distribution of critical points in its surroundings and their correspondence with the position of known clusters and groups in the area.

\begin{figure*}
\centering
\includegraphics[scale = 0.9]{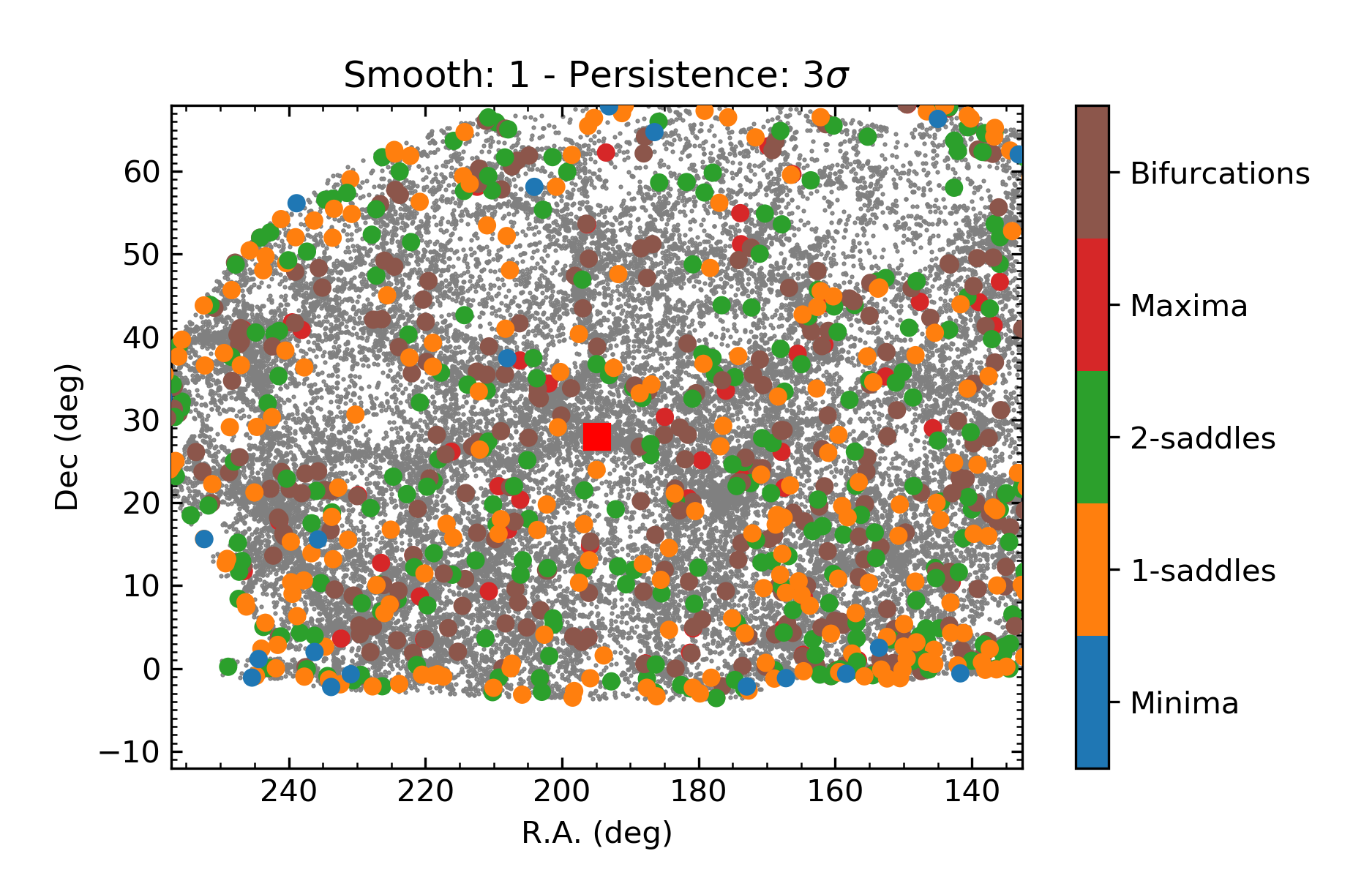}
\includegraphics[scale = 0.9]{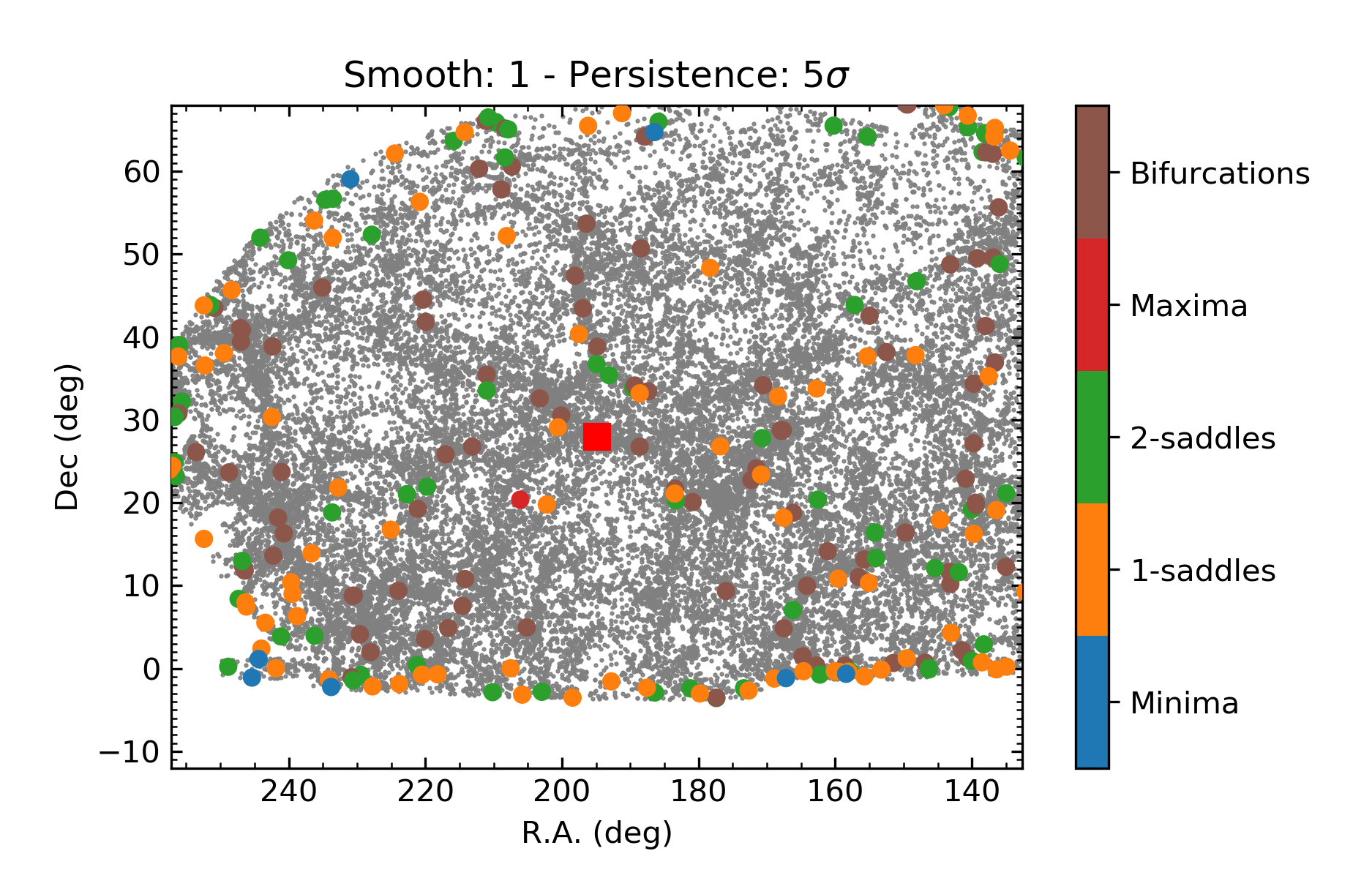}
\caption{Galaxy and critical point distribution in the Coma supercluster in a $\pm 75 \mathrm{Mpc}$ redshift slice centred on the Coma cluster (marked with the red square). Small grey points represent the position of galaxies from the MGS. Large circles represent critical points colour-coded according to their type. Top panel refers to a $3\sigma$ persistence threshold, bottom panel refers to a $5\sigma$ persistence threshold. 1 smoothing of the density field is applied prior to the filament and critical point detection. North is up, East to the left.}
\label{cp_in_coma_slice_smooth_1}
\end{figure*}

Figure \ref{cp_in_coma_slice_smooth_1} shows the critical points detected by \disperse (only those in a slice of $\pm 75 \mathrm{Mpc}$ centred on the Coma cluster are reported for clarity) superimposed to the galaxy distribution from the MGS in the same redshift interval. In the figure, both the $3\sigma$ and the $5\sigma$ persistence detection are shown for our fiducial LSS obtained with one iteration of smoothing of the density field.

In this figure, critical points are classified according to their type (maxima, minima, 1-saddles, 2-saddles, and bifurcations, i.e. critical points inserted where filaments intersect). The position of the Coma cluster is marked at the centre of the image. The distribution of critical points on the plane of the sky clearly follows the galaxy distribution. The majority of the points visible in the region are indeed 2-saddles (local density minima inside filaments) and bifurcation points, with very few minima. This is already an indication that the region around Coma is densely connected and that the Coma cluster occupies an important nodal position in the Cosmic Web. As expected, the number of critical points that remain visible after increasing the persistence threshold to $5\sigma$ drops, but still all the kinds of points, from bifurcations to minima, are present even when considering only the most secure ones. A $5\sigma$ persistence threshold may be rather high in that it selects only the most secure filaments and critical points, but it still offers a global view of the LSS around Coma. Appendix \ref{extrafigures} shows the changes happening to Figure \ref{cp_in_coma_slice_smooth_1} when a different level of smoothing is selected. As stated in Sect. \ref{method}, these figures show how in the no-smoothing case the number of detected critical points increases, with all the critical-point types being represented, while conversely in the 2-smoothing case very few absolute maxima and minima of the density field are detected.

Table \ref{crits_in_virial_radius} reports the number and type of critical points inside a sphere centred on Coma and defined by the virial radius of the cluster. We consider all these points as being associated with the galaxy over-density corresponding to Coma. This table shows how \disperse correctly identifies the Coma cluster with a maximum of the density field or with a point in space where several filaments intersect (bifurcation). The Coma supercluster system is never identified with a minimum or a saddle, except in the case of no smoothing and a $3\sigma$ persistence threshold. This is somewhat expected, since not performing any smoothing of the density field prior to the application of \disperse may result in the density field being more sensitive to shot noise. Substructures in the cluster and local variations of the density field, may indeed generate saddle points (i.e. local density minima) within the virial radius of Coma. It is also for this reason that we favour the \disperse run performed on the density field with one iteration of smoothing.

\begin{table}
\caption{Number of critical points inside the Coma virial radius. In each column the number refers to a $3\sigma$ persistence threshold, while the number in parentheses refers to a $5\sigma$ threshold.}
\label{crits_in_virial_radius}
\centering
\begin{tabular}{c c c c}
\hline\hline
Type & No smooth & 1-smooth & 2-smooth \\
\hline
Minima       & 0 (0) & 0 (0) & 0 (0) \\
1-saddles    & 0 (0) & 0 (0) & 0 (0) \\
2-saddles    & 1 (0) & 0 (0) & 0 (0) \\ 
Maxima       & 1 (0) & 1 (0) & 1 (0) \\ 
Bifurcations & 3 (1) & 1 (1) & 0 (1) \\
\hline
\end{tabular}
\end{table}

Using the critical points detected within the virial radius of the Coma cluster as an anchoring point, we recursively identify all the filaments connected to the Coma cluster up to a 75 comoving Mpc radius from the cluster position. Our approach consists in identifying the filaments connected to the critical points associated with Coma, to consider the critical points at the other end of those filaments and to identify all the filaments connected to them. This process is repeated until the 75 Mpc radius limit criterion is met.

\begin{figure*}
\centering
\includegraphics[scale = 0.9]{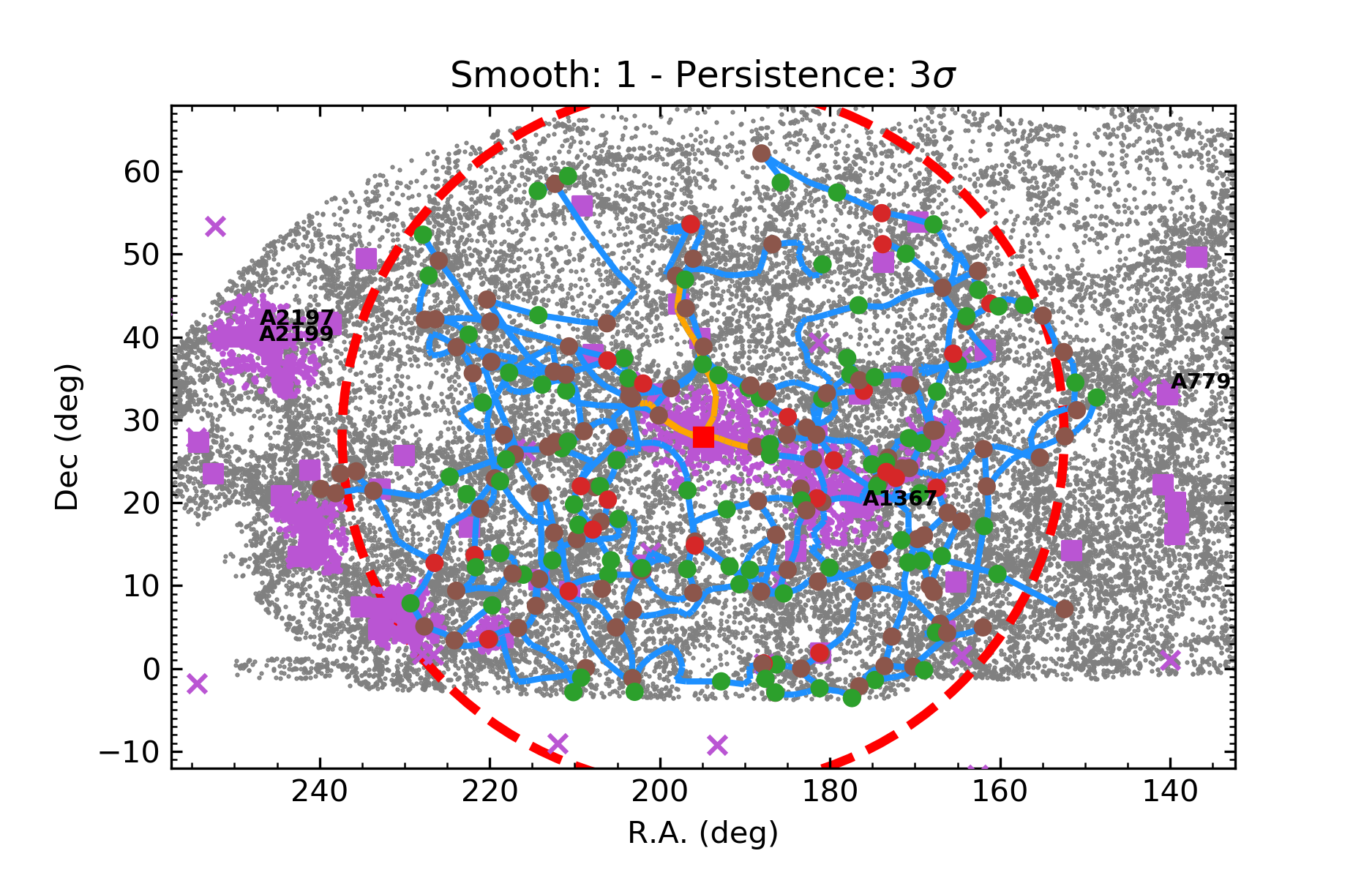}
\includegraphics[scale = 0.9]{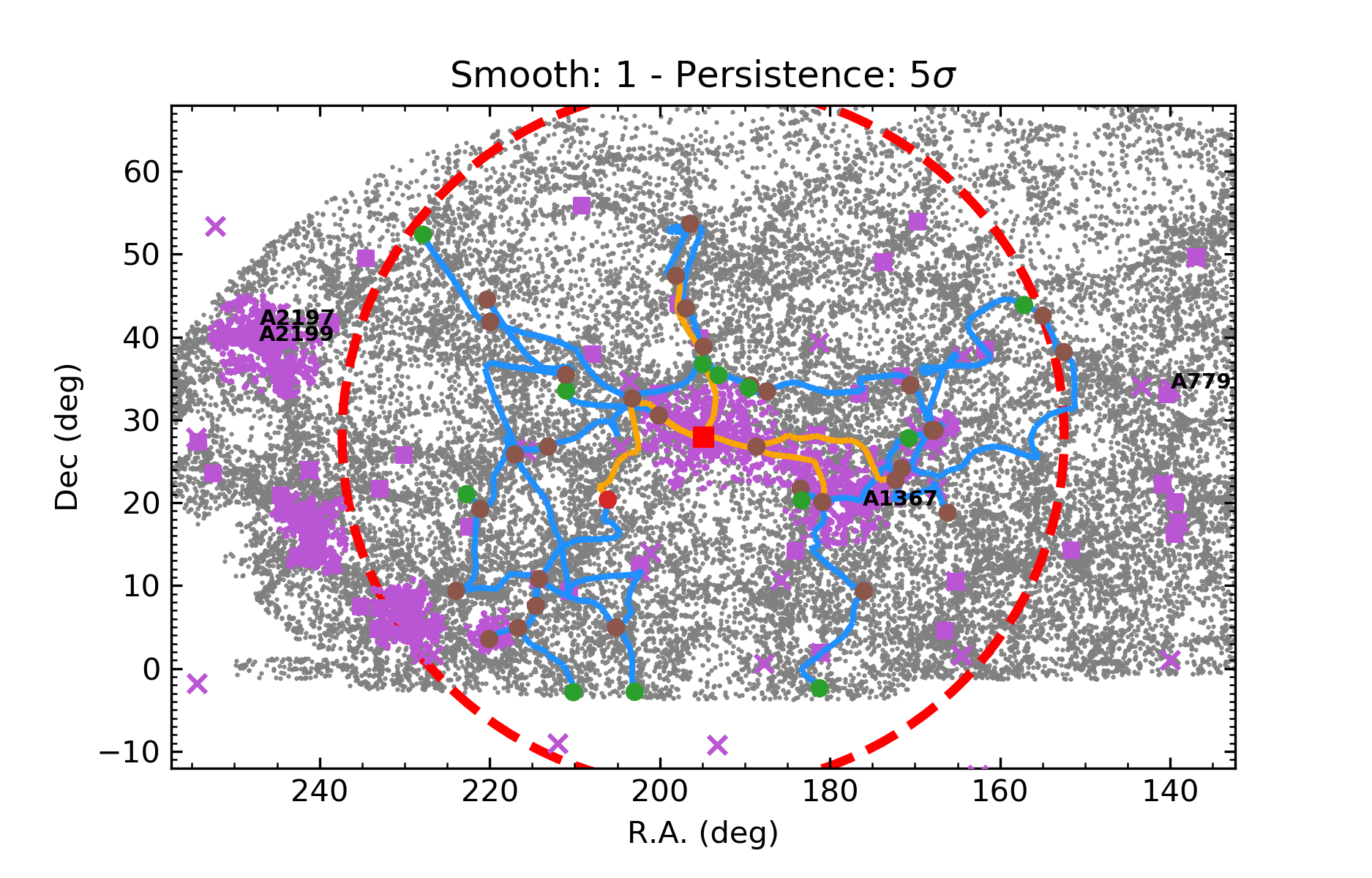}
\caption{Map of the filaments connected directly to the Coma cluster in a $\pm 75 \mathrm{Mpc}$ redshift slice centred on Coma. Small grey points represent the galaxies from the MGS. Large points represent critical points colour-coded according to their type. Cyan lines represent filaments connecting critical points as detected by \disperse. Orange lines are first and second generation filaments (see text). Purple crosses are MCXC clusters, purple squares are groups and clusters from the \citet{Tempel2017} sample (with $M_{200} \geq 10^{14} M_{\sun}$), small purple pentagons are SDSS supercluster member galaxies. The position of the Coma cluster is marked by a red square. The red dashed circle has a radius of 75 comoving Mpc and represents the projection of the sphere within which filaments connected to Coma were analysed. Top panel refers to a $3\sigma$ persistence and bottom panel to a $5\sigma$ persistence thresholds. North is up, East to the left.}
\label{fil_in_coma_sphere_smooth_1}
\end{figure*}

These filaments are shown in cyan in Fig. \ref{fil_in_coma_sphere_smooth_1}. As an example, in the same figure we show with a different colour (orange) the filaments directly connected to a critical point within the virial radius of Coma (which we refer to as ``first generation filaments'') and those directly connected to the critical points at the end of the first generation filaments and not within the Coma virial radius (``second generation filaments'').

Figure \ref{fil_in_coma_sphere_smooth_1} also shows the position of X-ray detected clusters from the MCXC sample, groups and clusters of galaxies from the \citet{Tempel2017} sample (only groups with $M_{200} \geq 10^{14} M_{\sun}$ are shown for clarity), as well as superclusters from the \citet{Liivamaagi2012} sample (shown in the figure through the position of their member galaxies to highlight their extended nature and complex morphology) which fall in the $\pm 75 \mathrm{Mpc}$ redshift slice centred on Coma.

Visually, the filaments connected to the Coma cluster seem to closely follow projected over-densities of galaxies, often passing through the position of known clusters and groups, while several critical points are found at a close distance from cluster positions. In particular, filaments connect Coma to the position of Abell 1367 (identified in the \citealt{Liivamaagi2012} sample as a single structure with Coma) and depart from Coma in the direction of Abell 779 and Abell 2199/2197.

\subsection{Cluster matches and filament bundles}
\label{critmatches_filbeams}
In order to quantify the connection of the Coma cluster to other structures in the region, we performed a match between the positions of critical points and filaments as detected by \disperse and those of already known clusters from the literature, namely the samples introduced in Section \ref{data_cluster}.

We define critical points as matched to clusters, if the points are lying within a sphere of radius $R_{500}$ ($R_{200}$ in the case of \citealt{Tempel2017} optical groups) centred on the cluster position. The same applies to filaments, which are considered as matched to a cluster if at least one segment has its midpoint within a sphere of radius $R_{500}$ ($R_{200}$) centred on the cluster position.

\begin{figure}
\includegraphics[scale = 0.725, trim = 0cm 2cm 0cm 2cm, clip = true]{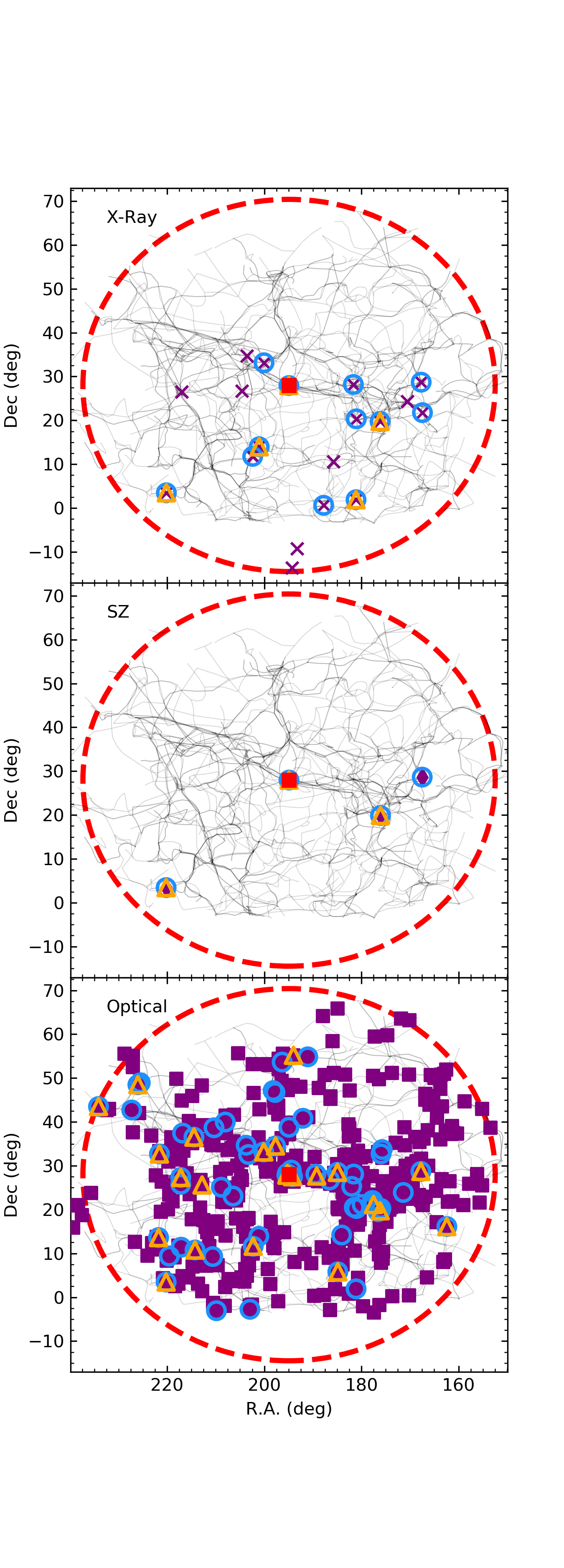}
\caption{Matches between clusters from the literature (MCXC: top, Planck clusters: middle, SDSS optical groups from \citealt{Tempel2017}: bottom) and critical points and filaments. Purple crosses, diamonds, and squares mark the position of all the clusters/groups within a 75 Mpc sphere centred on the Coma cluster (whose projection on the plane of the sky is marked by the red dashed circle). The position of the Coma cluster is marked by the red square at the centre. Cyan circles mark those clusters/groups which have a match with at least one filament, while orange triangles mark those clusters which have a match with at least one critical point, regardless of smoothing and persistence threshold. Black lines in the background show the position of all filaments regardless of smoothing level or persistence threshold. In each panel North is up, East to the left.}
\label{critmatch}
\end{figure}

Figure \ref{critmatch} shows the matches between critical points and filaments with clusters and groups from the literature used for our analysis and described in Section \ref{data_cluster}. In this figure, all the clusters within the 75 Mpc radius sphere around Coma are reported, with those matched to critical points highlighted by orange triangles and those matched to filaments highlighted by cyan circles. In the background of each panel, all filaments from all persistence thresholds and smoothing levels are plotted in black. This includes all filaments from Fig. \ref{fil_in_coma_sphere_smooth_1} as well as Fig. \ref{fil_in_coma_sphere_no_smooth} and Fig. \ref{fil_in_coma_sphere_smooth_2}.

A large number of the clusters present within the 75 Mpc radius region of interest is matched to either a filament or a critical point, with the former occurrence happening more often than the latter. Although the \disperse algorithm introduces some degree of uncertainty, due to its dependence on the arbitrarily selected degree of smoothing and persistence threshold, still clusters are systematically matched to critical points and filaments in regions of space where several filaments from the various smoothing and persistence combinations overlap. This nicely illustrates how clusters are indeed generally located at the intersection of filaments (nodes) as well as found along the filaments (knots).

\begin{figure}
\includegraphics[width=\linewidth]{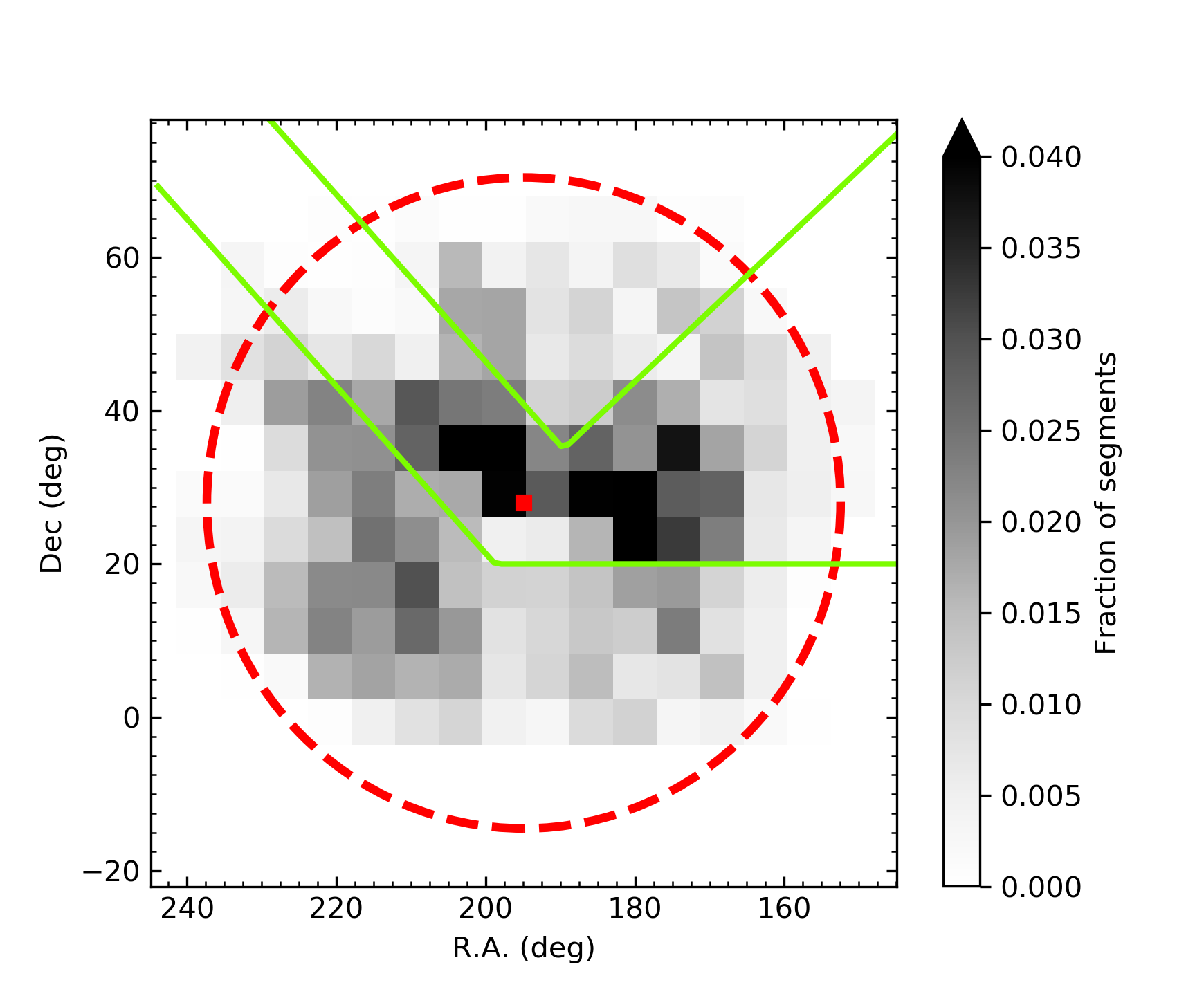}
\caption{Map of the filament bundles in the Coma cluster region. The grayscale histogram shows the fraction of filaments that cross a given pixel regardless of persistence threshold and smoothing level. The red dashed circle has a radius of 75 comoving Mpc and shows the projection of the sphere around Coma which we considered for our analysis. The position of the Coma cluster is marked by the red square. Green lines show the region inside which we derived some of the profiles described in Sect. \ref{results_gasphase}. North is up, East to the left.}
\label{filbeams}
\end{figure}

We note how along the paths marked by the cluster positions, filaments seem to define bundles that depart from the Coma cluster. This indicates that all the combinations of smoothing and persistence cuts identify the same overall large scale structures. This is made even more evident in Fig. \ref{filbeams}, which presents a smoothed version of the background of Fig. \ref{critmatch}. Figure \ref{filbeams} shows a two-dimensional histogram, where the intensity of the $10 \times 10$ Mpc pixels is proportional to the fraction of filament segments from all the persistence and smoothing combinations crossing them.

This figure clearly shows how there are at least two bundles of filaments directly connected to Coma: one extends North-East of the cluster, at roughly a $45\deg$ angle, while the other starts South-West of the cluster, to then bend directly West (highlighted by green lines in the figure). More detail on the filaments close to the Coma cluster will be given in Section \ref{discussion}.

\subsection{The connectivity of the Coma cluster}
\label{connectivity_section}
Connectivity is a measurement of the number of filaments directly connected to a node of the Cosmic Web. It is an important way of providing a quantitative analysis of the position of a cluster in its surrounding LSS. Moreover, as the Cosmic Web evolution tends to disconnect structures and merge filaments together (although this is debated, see e.g. \citealt{Codis2018}, \citealt{DarraghFord2019} and references therein), measuring connectivity for clusters of galaxies as a function of mass and redshift can provide a test for theoretical models of LSS formation. The expected trend is for cluster connectivity to increase as a function of mass, with more massive clusters being connected with up to five filaments \citep[see e.g.][]{Pimbblet2004, Colberg2005, AragonCalvo2010, Sarron2019, DarraghFord2019}. 

We measure the connectivity of the Coma cluster as the number of filaments which have one extremum inside the sphere defined by the Virial radius of Coma and the other extreme located at a distance from Coma greater that 1.5 times the Virial radius. According to this criterion, we find that Coma is connected to between 3 and 2 filaments (depending on smoothing levels and persistence thresholds), with a median connectivity of 2.5.

\begin{figure}
\includegraphics[width=\linewidth]{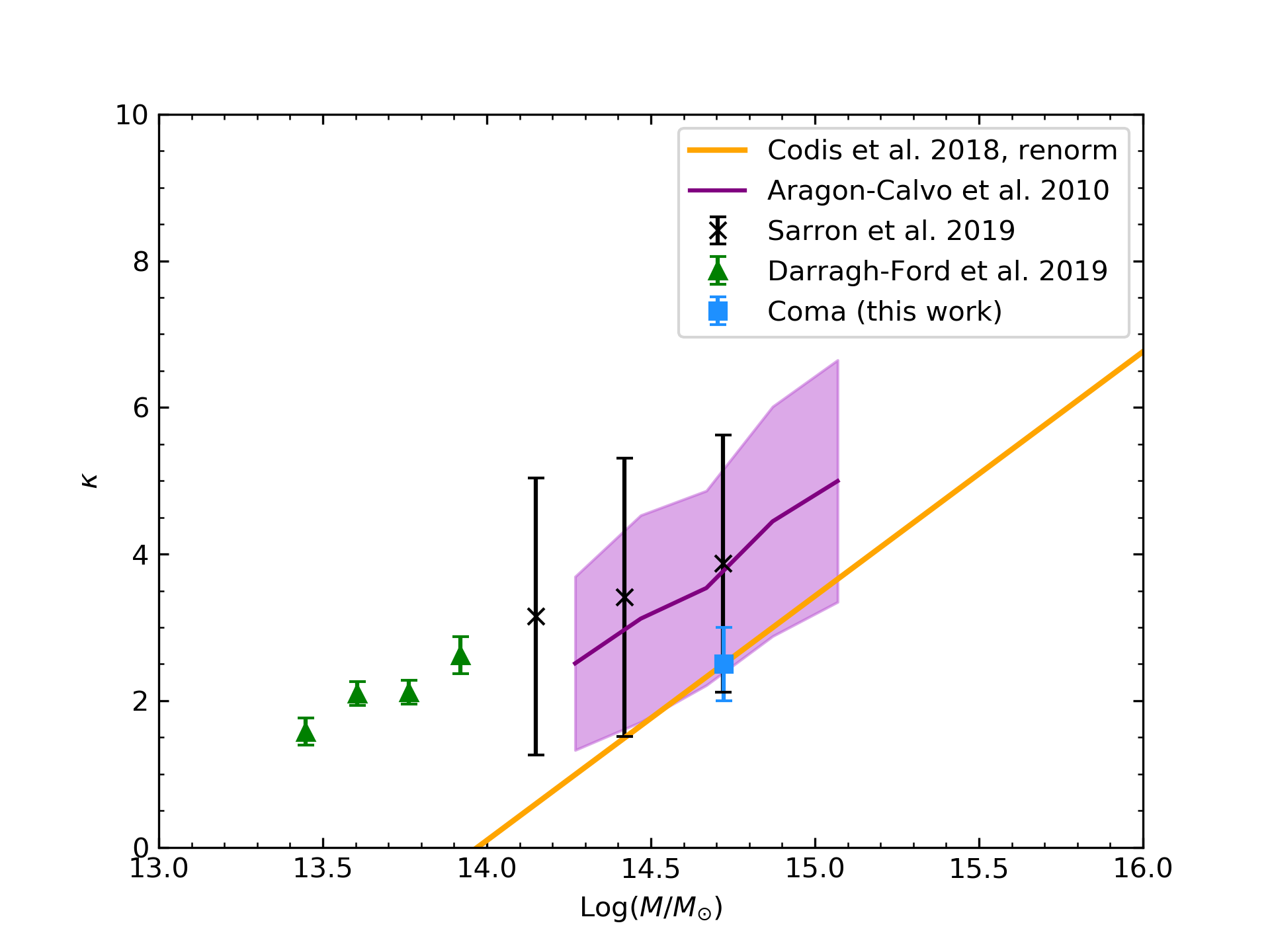}
\caption{Connectivity of the Coma cluster. The cyan square refers to the Coma cluster as analysed in this work. Black crosses are the observed connectivity values from the AMASCFI Clusters \citep{Sarron2019}, green triangles are the observed connectivity values for the groups in COSMOS \citep{DarraghFord2019}, the purple line and shaded region are the relation from the numerical N-body simulations of \citet{AragonCalvo2010}, and its corresponding 1$\sigma$ uncertainty, while the solid orange line is the theoretical relation of \citet{Codis2018} renormalised to pass through our measurement for the Coma cluster, so as to better compare trends. Note that the mass on the $x$-axis of the plot is $M_{200}$ for the Coma, \citet{DarraghFord2019}, and \citet{Sarron2019} points and a value close to $M_{vir}$ for the \citet{AragonCalvo2010} and \citet{Codis2018} relations.}
\label{coma_connections}
\end{figure}

Figure \ref{coma_connections} shows the measurement of the connectivity for the Coma cluster, compared to similar measurements from the literature derived both from observations and theory. The trend of a connectivity increase with mass is present in several measurement from the literature and it is consistent from low-mass groups \citep{DarraghFord2019} to massive clusters \citep{Sarron2019}.

The connectivity measurement for Coma is in broad agreement with connectivity measurements from the literature. For example, \citet{Sarron2019} reported measurements of the connectivity from AMASCFI clusters \citep{Sarron2018} to be in the range $\kappa = 3 \div 4$ in the mass range $\log(M_{200}/M_{\sun}) \gtrsim 14$ (black crosses in Fig. \ref{coma_connections}). This value is slightly larger than our estimate for Coma, but the difference could be due to the connectivity of the \citet{Sarron2019} clusters being measured as the number of filaments crossing a sphere of radius 1.5 comoving Mpc centred on the clusters, i.e. on a much smaller scale than what we use for our measurement. In the Coma cluster case, this sphere would be completely located inside the virial radius and it is possible that this is the reason for the slightly higher connectivity measurement reported by \citet{Sarron2019}. Indeed, if we measured the connectivity for the Coma cluster on the same scale as \citet{Sarron2019}, we would obtain a measurement in the range $\kappa = 3 \div 9$.

From the theoretical point of view, we compare our results with \citet{AragonCalvo2010}, who measured the connectivity of clusters detected using the Multi-scale Morphology Filter \citep[MMF,][]{AragonCalvo2007} in a N-body simulation. Their relation shows that the connectivity increases with mass \citep[in agreement with][]{Sarron2019} ranging between $\kappa = 2 \div 5$ in the mass range $\log(M/M_{\sun}) \sim 14 \div 15$. Although in agreement with our estimate, still our measurement and the one by \citet{AragonCalvo2010} are not completely comparable: their mass estimate as derived from the MMF formalism is closer to $M_{vir}$ rather than $M_{200}$, while we use $M_{200}$ for our measurement to be consistent with \citep{Sarron2019}.

Figure \ref{coma_connections} also reports the analytic relation derived by \citet{Codis2018}. In this work, the authors derived a functional form for the relation between cluster/node mass and connectivity. As they derived this relation on a dark matter only simulation, with the \disperse algorithm run on dark matter halos derived with a Friend-of-Friend method on the full particle distribution, the relation may not be fully representative of a reality, where under-sampling due to observational issues may be present. For this reason, the normalisation of the relation may be biased high, as all the filaments will be recovered in the simulations. Still, we report the relation in the same figure with our measurement, renormalised so as to pass through the value for the Coma cluster. The shape of the relation by \citet{Codis2018} is found to be in very good agreement with what found by \citet{AragonCalvo2010}. On the other hand, the measurements by \citet{Sarron2019} seem to be flatter than the relation by \citet{Codis2018} at low cluster masses. This trend persists when extending the comparison to lower mass groups such as those detected in COSMOS \citep{DarraghFord2019}, which show considerable deviation from the renormalised \citet{Codis2018} relation. Nevertheless, the error bars on the \citet{DarraghFord2019} points only represent the error on the mean and not the standard deviation (as is the case e.g. for \citealt{Sarron2019}). Therefore the agreement may improve if more realistic error bars were considered.

\section{Filament detection in the SZ}
\label{results_gasphase}
Having determined a reliable sample of filaments (or filament bundles) in a 75 comoving Mpc radius sphere around Coma, we decided to investigate a possible detection of these filaments in the SZ signal.

We start by analysing the signal in the filament bundles. We identify in Fig. \ref{filbeams} all the pixels crossed by a fraction of segments above the 75th percentile of the distribution  of fraction values (hereafter: ``inside bundles'') as well as the complementary sample of pixels below the 75th percentile of the distribution of fraction values (hereafter: ``outside bundles''). We sum and average the SZ signal in the pixels excluding masked regions according to Sect. \ref{planck_map_data}.

In order not to include in our analysis the signal from the Coma cluster, we subtracted a 2D profile from the \emph{Planck} $y$-map for the cluster SZ signal itself. We modelled the Coma SZ signal with the \citet{Arnaud2010} Universal Profile, using the parameters by \citet[][Table 1, Model C: ``Universal'' all free]{PlanckCollaborationXComa}. The 3D profile is projected on the plane of the sky at the position of the Coma cluster (using the SZ centroid coordinates, see Table \ref{coma_values}), and it is matched to the \emph{Planck} $y$-map resolution by convolving it with a Gaussian of $\sigma = 10\arcmin$. The modelled cluster SZ signal is then subtracted from the maps out to a distance of $10 \times R_{500}$ from the position of Coma prior to further analysis. This is done in order to recover the SZ signal from the filaments (including in regions close to the cluster where significant deviations from a spherical symmetry may indicate the connection to the large scale structure, similar to what done in the X-rays when performing fluctuation analysis, see e.g. \citealt{Neumann2003}).

The total SZ signal inside bundles is $y_{inside} = -0.23 \pm 0.03$ while outside is $y_{outside} = -0.45 \pm 0.04$. This yields a ratio of the signal $y_{inside}/y_{outside} = 0.50$ with a ratio of the number of pixels $N_{inside}/N_{outside} = 0.34$. Error bars on the measurements have been derived through jackknife resampling, by systematically excluding one of the pixels, repeating the analysis, and considering the variance of the resulting measurements distribution.

These values seem to indicate an excess of signal in the regions of space where filament bundles are located, consistent with a tentative detection of the filaments in the gas phase. If we consider the average of the signal inside the bundles and outside, we obtain a value of $\bar{y}_{inside} = -2.20 \times 10^{-7} \pm -2.58 \times 10^{-8}$ inside the bundles and a value $\bar{y}_{outside} = -1.56 \times 10^{-7} \pm -1.59 \times 10^{-8}$ outside the bundles. The significance of the detection is therefore $2.1\sigma$.

We also investigate the possibility of deriving the profiles of the filaments around the Coma cluster, as done in Tanimura et al. (in preparation) for the full sample of filaments in the SDSS in the SZ and by Bonjean et al. (in preparation) for the galaxy distribution.

Following Bonjean et al. (in preparation), we made use of a modified version of the public software RadFil\footnote{\url{https://github.com/catherinezucker/radfil}, modifications by V. Bonjean, private communication.} \citep{Zucker2018} to derive the filament profiles. RadFil measures the profile of pixel intensity on pixelised maps, in line-of-sights perpendicular to a given spine path. Our input to the code consists of cut-outs of the \emph{Planck} $y$-maps of $20 \times 20 \deg$ (to reduce distortions due to flat-sky approximation at the map edges) which follow each filament in the 75 comoving Mpc radius region of interest (cyan lines in Figure \ref{fil_in_coma_sphere_smooth_1}) and the set of segments constituting each individual filament.

The result is a set of profiles, on perpendicular lines spaced by one pixel along the spine of the filament. We collapse these profiles on the direction along the filament, averaging them and obtaining a single profile per filament. We repeat the procedure for each filament in the 75 comoving Mpc radius region of interest around Coma, examples of the map patches for individual filaments and of the resulting one pixel-spaced and averaged, collapsed profiles are presented in Appendix \ref{prof_and_maps}.

\begin{figure}
\includegraphics[width=\linewidth]{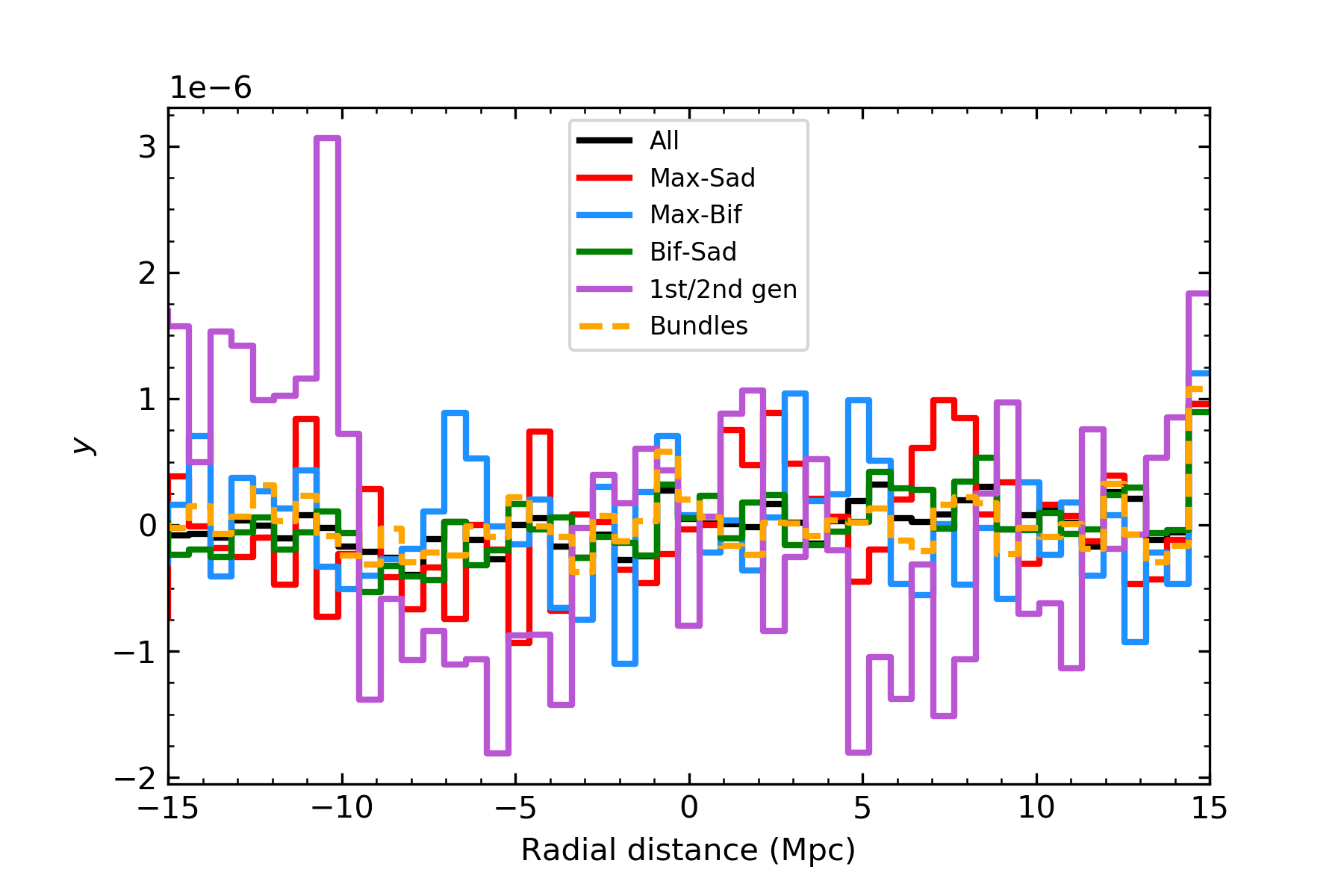}
\caption{Filament profiles in the SZ. The black line is the average of the collapsed profiles for all filaments in the Coma region. Cyan, red and green lines are the average of the collapsed profiles for filaments between maxima and saddles, maxima and bifurcations, and bifurcations and saddles, respectively. The purple line is the profile for the first and second generation filaments, while the orange dashed line is the profile for the filaments in beams. Each profile has been normalised to its own individual mean.}
\label{prof_all}
\end{figure}

As the SZ signal is too faint to obtain a detection of each single filament in the gas phase, we averaged the collapsed profiles for all the filaments in the region as well as for different subsamples of filaments, such as those with different types of critical points at their extrema (maximum-saddle, saddle-bifurcation, maximum-bifurcation), first and second generation filaments as defined in Sect. \ref{results_connections} (orange lines in Fig. \ref{fil_in_coma_sphere_smooth_1}), and filaments inside bundle regions as defined by the set of green lines presented in Fig. \ref{filbeams}. The average of the collapsed profiles for all these categories of filaments is visible in Fig. \ref{prof_all}.

The SZ profiles look rather flat and noisy, without particular features in the $\pm 15$ Mpc range from the filament axis. Our conclusion is that we are able to detect the total signal in the SZ from filament bundles in the 75 Mpc region around Coma, but not to characterise the filament profiles in the gas phase. This is reasonably due to a statistics problem, as increasing the filament sample to include all filaments in the SDSS a detection is possible (Tanimura et al., in preparation). In conclusion, due to the small number of filaments analysed here, we are unable to detect a signal that we expect to be smaller than the $y$-map noise. As an example, by stacking $\sim 260\,000$ pairs of Luminous Red Galaxies in the SDSS DR12, \citet{Tanimura2019} detected a stacked signal in the SZ for the filaments between these galaxies of $\Delta y \sim 10^{-8}$ at $5.3\sigma$. As in our analysis of the filament profiles we are dealing with (depending on the sample of filaments considered in Figure \ref{prof_all}) between $\sim 7 \div 260$ filaments, we expect a signal a factor $10^{5}$ smaller.

\section{Discussion}
\label{discussion}
The goal of this paper is to characterise in a quantitative way the LSS around one of the best known galaxy clusters. The Coma cluster is a massive one, with a high degree of substructure and several infalling groups along directions which seem to connect it to other galaxy clusters. We confirm this picture and quantify it through the use of the \disperse$\:$ algorithm.

\subsection{Indirect confirmation of the first generation filaments}
Coma is connected to at least three secure filaments, which connect it to other clusters in the region, and several other critical points or filaments connected to Coma are well matched or intersect the position of clusters detected in the optical, X-ray, and from \emph{Planck}. The three first and second generation filaments (i.e. directly connected to a critical point inside the virial sphere of Coma) are directed along the NE-SW axis (with Coma at the SW end), along the N-S axis (with Coma at the southern end) and along the E-W axis (with Coma at the eastern end) when we consider them counter-clockwise starting from the west in Fig. \ref{fil_in_coma_sphere_smooth_1}. A hint of the presence of the three secure filaments detected in our study can also be found in \citet{Mahajan2018} where the authors analysed a smaller patch centred on the Coma cluster and Abell 1367. Their analysis showed threads of filament galaxies departing from Coma to the North, to the West (connecting Coma to Abell 1367) and to the North-East of the cluster. This focus on the intermediate/small scale of the Coma-Abell 1367 system complements our analysis of the LSS around Coma on a much larger scale. Looking more in detail at the properties of the three secure filaments we have detected and where they connect to the Coma cluster, we find a very good agreement with the position of known features of the cluster emerging from X-ray, SZ and optical analyses in the literature \citep{Neumann2003, Adami2005ICL, PlanckCollaborationXComa, Lyskova2019}.

In particular \citet{Neumann2003} analysed \emph{XMM-Newton} data in the Coma region and produced a map of the residuals left in the X-ray distribution after the subtraction of a best-fit $\beta$-model. This residual map shows a feature directly to the West of Coma aligned in the N-S direction, curved along the cluster edge, with a temperature increase from 8 keV to 10 keV between one side of the structure to the other \citep[see e.g. Figures 2 and 3 of][]{Neumann2003}. Such a sharp temperature transition across a structure detected in the X-ray can be interpreted as the feature being a shock front due to matter accretion on the cluster. The position of this shock arc is perfectly consistent with the point where our western filament (aligned in the E-W direction and therefore perpendicular to the shock front direction) connects to the cluster. We therefore express the possibility that the shock front is due to the accretion of cosmic matter coming from our identified filament. The fact that this filament is also identified in our 2D analysis (done to check for the effect of Finger of God redshift distortions, see \citealt{Kaiser1987}, on our Cosmic Web reconstruction, see Appendix \ref{fog}) approximately in the same position confirms our conclusion that the filament is a real structure connecting to Coma.

On the same side of the Coma cluster, southern of the \citet{Neumann2003} shock, is located the NGC4839 group. This galaxy and the associated group present a well detected feature both in the X-rays and in the SZ signal. The most common explanation for the presence of this feature is that these objects are infalling on the cluster for the first time from a filament located in the direction of Abell 1367, SW of the Coma cluster \citep[see e.g.][]{Neumann2001, Neumann2003, Adami2005, Adami2007LF, BrownRudnick2011, OgreanBruggen2013}. We do not detect the presence of such a filament in the \disperse Cosmic Web. While there is indeed a filament connecting to Coma in the direction of Abell 1367, it is the aforementioned E-W aligned filament, with no structure connecting to the cluster at the SW corner where NGC 4839 is located. Moreover, no combination of \disperse parameters seem to recover such a filament (see the background black lines corresponding to the filament bundles in Figure \ref{critmatch}). Although this finding is in clear disagreement with the scenario of first infall for the NGC 4839 group, it is actually very consistent with recently proposed analyses \citep[e.g.][]{Sheardown2019, Lyskova2019} which advocate for a post-merger scenario for NGC 4839, where the galaxy has crossed the cluster core already once and is now coming out the other side. Quite interestingly, in this case the infall direction for NGC4839 would be the NE corner of the cluster, where \disperse finds a quite significant filament. This filament is our eastern filament, aligned in the NE-SW direction, the same one showed by the NGC 4839 X-ray and SZ tail. If confirmed, this scenario would potentially identify in the NE-SW filament with Coma at the SW end the filament where NGC 4839 came from. Also this filament is present in our 2D analysis, therefore increasing the chance of it being a real structure. The idea of the NE direction as a possible one for infall has also been supported by \citet{Adami2007LF}, when discussing the finding of a more populated luminosity function in the NE region with respect to the SW one.

As for the northern filament, aligned in the N-S direction with Coma at the southern end, this filament does not match any clear feature from the X-rays or SZ analyses. It is also not present in our 2D analysis, as well as missing from some other combinations of \disperse parameters, making it the least secure of the first-generation filaments. Still, it is present in the skeleton recovered with a $5\sigma$ persistence threshold in the 1-smooth and no-smooth cases, an indication of it being rather significant. In their analysis of the intra-cluster light (ICL) in the central region of the Coma cluster, \citet{Adami2005ICL} found a possible source of ICL north of the two central dominant galaxies of the cluster (e.g. Source 4 in \citealt{Adami2005ICL}). With a possible origin for ICL commonly indicated in the disruption of accreted galaxies, together with the fact  that no sources of ICL seem to be identified in the southern area of the cluster, this is a tempting indication of the presence of an accreting cosmic filament connecting to Coma in the northern region. Nevertheless, the small region of focus of the \citet{Adami2005ICL} analysis around the two central dominant galaxies compared to the LSS analysed in this work, as well as the high degree of possible systematics and contaminating sources to the ICL combined with the intrinsic uncertainties of the \disperse method, prevent us from making any further connection between the presence of ICL and that of accreting filaments.

As a final remark, \citet{Adami2009PFA} performed an analysis of the environment along the Line-of-Sight (LoS) of Coma, using very deep spectroscopy of faint galaxies up to $z \lesssim 0.2$. By applying an algorithm to detect groups starting from galaxy positions and their luminosities, they detected the hint of a structure (called the Putative Filament Area, PFA), traced by several galaxy over-densities (knots), connecting Coma with another structure at redshift $z = 0.054$ (the Background Massive Group, BMG). As part of this structure should be within our 75 Mpc radius region of interest around Coma, we checked in the filaments recovered by \disperse whether we find a match with the PFA. Our analysis of the filaments so close to the Coma cluster is complicated by the Finger of God redshift distortions along the LoS. We do however have a filament along the Line-of-Sight, connecting to Coma approximately at the same position and along the same direction as the PFA structure. However, the PFA detected by \citet{Adami2009PFA} is defined on a much smaller scale than what analysed here and the position and direction of the filament is only hinted at by the position of knots detected on the LoS. This prevents us from drawing any further conclusion on the match between the filaments from \disperse and what found by \citet{Adami2009PFA}.

\subsection{A general picture}
Coma is a good example of a massive, highly connected cluster. These kind of objects, when found in numerical simulations, are invariably connected to several filamentary structures and occupy important positions in the Cosmic Web as high density peaks. For example, \citet{AragonCalvo2010} identified in N-body simulations massive $10^{14} M_{\sun}$ halos at the centre of complex networks of branches. Moreover, when identifying LSS features with the MMF, they found a class of ``star'' filaments, often hosting a cluster at their centre with several branches departing from it. This image well corresponds to what we find for the Coma cluster. In almost all combinations of smoothing and persistence for the \disperse algorithm, the cluster is identified with a maximum of the density field, or with a bifurcation point where several filaments are crossing.

From a more quantitative perspective, the connectivity of the cluster is between 2 and 3 (taking the effect of the combination of \disperse parameters into account). This value is in line with similar measurements from data in the literature and with what expected from numerical N-body simulations. Still, albeit consistent and with all the possible biases present in this comparison identified in Sect. \ref{connectivity_section}, Coma connectivity may be slightly too low with respect to what expected (e.g. by numerical simulations), for clusters of the same mass. If confirmed beyond the uncertainty intrinsic to the \disperse method and the possible biases in the comparison, this could help to better understand what happens to the cluster connectivity when mergers or significant infall of matter is ongoing.

Indeed, two of the filaments connecting to Coma (the eastern, NE-SW aligned one and the western, E-W aligned one) are consistent with being the location of significant accretion onto the cluster. This is in agreement with the picture of matter infalling onto clusters through the filamentary network \citep[see e.g.][and references therein]{Cautun2014}. Significant infall through filaments and merging activity with groups can in principle change the connectivity of a cluster, increasing it by having the resulting halo connected to the filaments of both the halos that have merged, or decreasing it by disconnecting halos from the Cosmic Web. Studies of the connectivity such as those presented in this work, may help in better understanding these processes.

The power of the \disperse method applied to galaxy surveys is also that of providing a rather secure indication of where filaments are located in a region around the clusters. This information can be exploited to have a characterisation of the filaments in the gas phase. While the detection of filamentary structures from the galaxy distribution is now starting to become commonplace in the literature, the detection of these objects from the gas phase is still largely missing, aside from known individual objects such as bridges (e.g. the one between Abell 399 and Abell 401, see \citealt{Bonjean2018}) or through stacking analyses \citep[e.g.][]{Tanimura2019, Tanimura2019Superclusters}.

In this work, we are able to detect the total signal of the filaments around Coma, in the regions of space where the presence of filament bundles is confirmed. This detection (if confirmed to a higher $\sigma$) would be on a much larger scale than the filaments detected in the X-ray close to the cluster core by \citet{Sanders2013}. In fact the kind of structures that we are investigating would be at the same scale than the bridges of matter detected between galaxy clusters by e.g. \citet{Akamatsu2017, Bonjean2018}, but of a different nature, the filaments between Coma and other clusters being longer and possibly thinner.

\section{Conclusions and summary}
\label{conclusions}
In this work we ran the \disperse algorithm on the SDSS DR7 MGS and analysed the resulting LSS distribution around the Coma cluster. By studying the number of filaments connected to Coma and their matching to the position of known clusters in the region, we found that:

\begin{enumerate}
\item Coma is situated in a densely connected network of filaments, potentially connecting it to several other clusters.
\item At least three secure filaments depart from the Coma cluster, in the already known direction of infalling groups as determined by other works, one to the West (whose position is consistent with a shock detected in the X-rays), one to the North and one to the North-East, whose position is consistent with the direction of infall of the galaxy NGC4839 and its associated group.
\item The connectivity value for Coma is in good agreement with other measurements for clusters of similar mass and with what derived from N-body numerical simulations, still with the possibility of it being slightly too low. 
\item An analysis of the \emph{Planck} $y$-map at the position of the filaments yields a $2.1\sigma$ significance detection of the total signal from the filaments in the region, but the signal-to-noise ratio is too low to derive their profile. Still we provide a tentative detection of the filamentary structure traced by \disperse around Coma in the gas phase.
\end{enumerate}

This work presents a case study of the properties of the Cosmic Web around a massive cluster. Although the LSS can be detected rather easily from the galaxy distribution, a complete characterisation in terms of the connections among its components and of the study of the filaments in terms of their galaxy and gas distribution is only recently starting to be achieved.  Filaments are the preferred direction for matter infall on clusters, they are the pathway for galaxies to be accreted onto massive halos and can play a role in preprocessing galaxies and turn them quiescent before they reach clusters, with important implications for the study of galaxy evolution. On the other hand, the number of filaments connected to a cluster may change with cosmic time and cluster evolution, therefore providing implications for the study of cosmology and structure formation. Studies of the Cosmic Web like the one presented in this paper, are crucial as they provide a complete picture of the position occupied by a cluster as a node of the Cosmic Web and present many implications for the study of these objects.

\appendix
\section{Exploring the \disperse parameters}
\label{extrafigures}
The \disperse algorithm offers the possibility for the user to choose the value of different parameters for the filament detection. The two main one to which the LSS is more sensitive are the smoothing of the density field and the persistence threshold. We present below different versions of Figures \ref{cp_in_coma_slice_smooth_1} and \ref{fil_in_coma_sphere_smooth_1} obtained with different levels of smoothing of the density field and different persistence thresholds. Increasing the smoothing reduces the number of critical points detected as it reduces the intrinsic Poisson noise of the density field. On the other hand, increasing the persistence threshold removes low significance critical points, therefore also reducing the number of critical points available in the final catalogue. In both situations, increasing the smoothing and/or the persistence threshold removes short filaments and increases the length of those remaining in the final detection.

\begin{figure*}
\centering
\includegraphics[scale = 0.9]{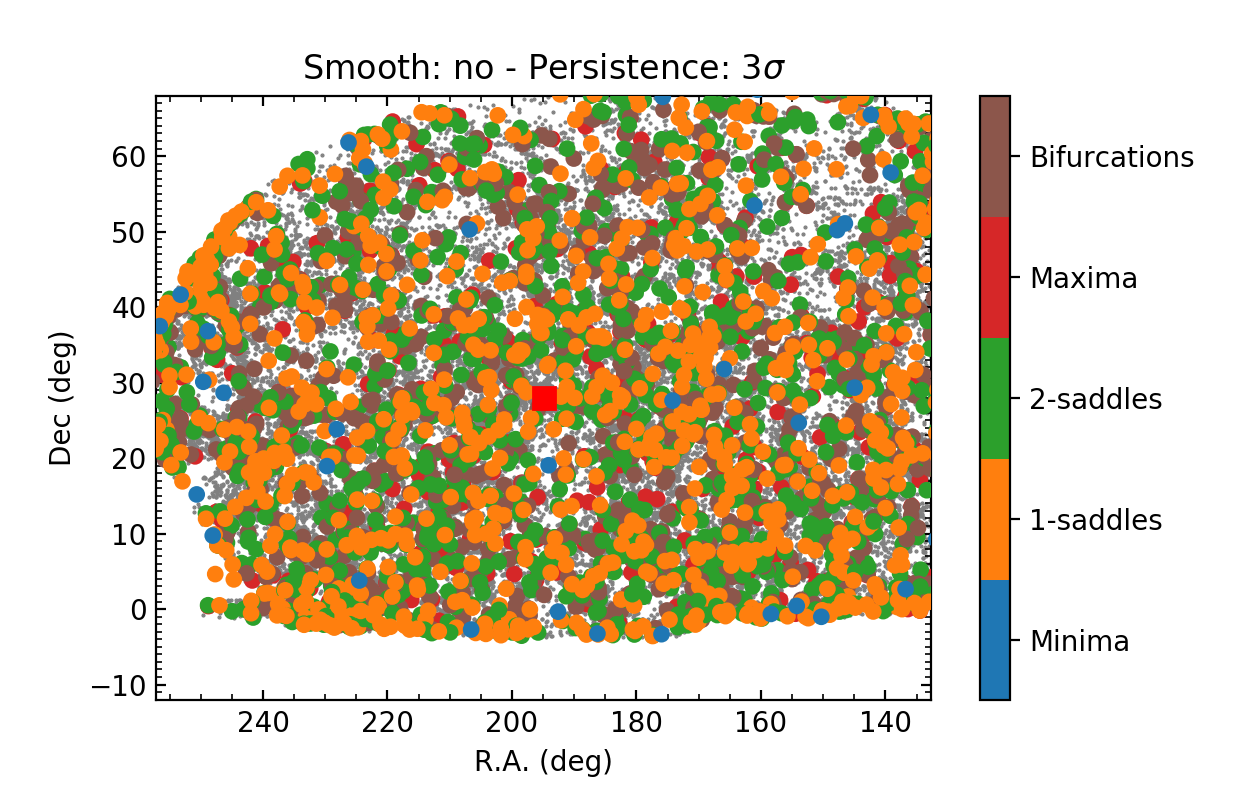}
\includegraphics[scale = 0.9]{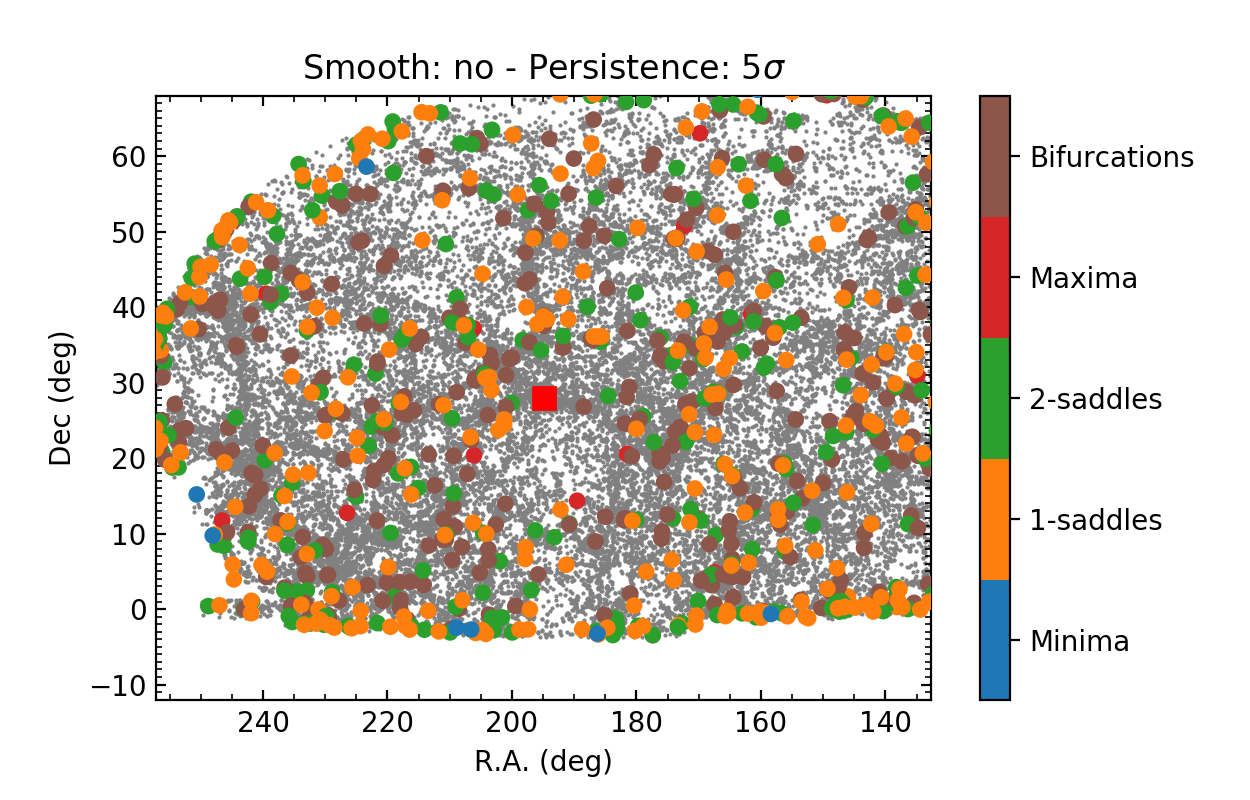}
\caption{Same as Figure \ref{cp_in_coma_slice_smooth_1}. No smooth of the density field.}
\label{cp_in_coma_slice_smooth_no}
\end{figure*}

\begin{figure*}
\centering
\includegraphics[scale = 0.9]{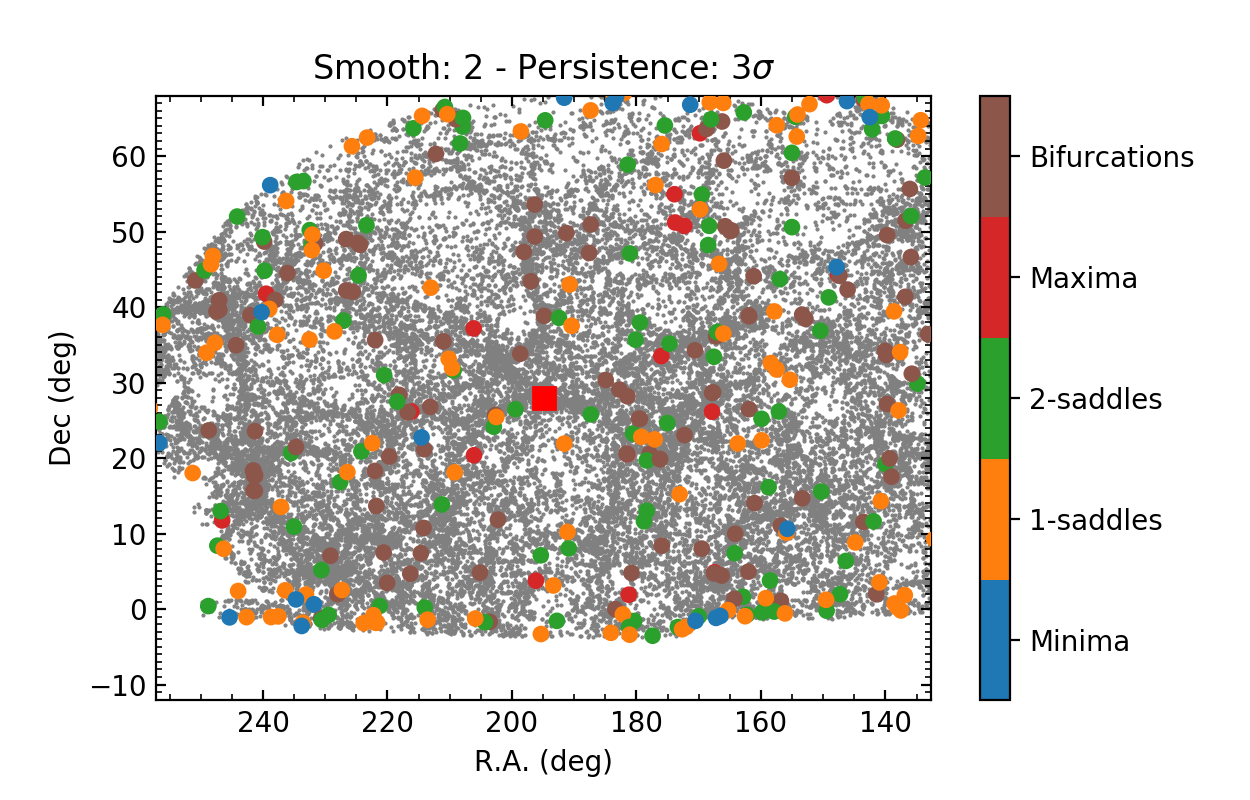}
\includegraphics[scale = 0.9]{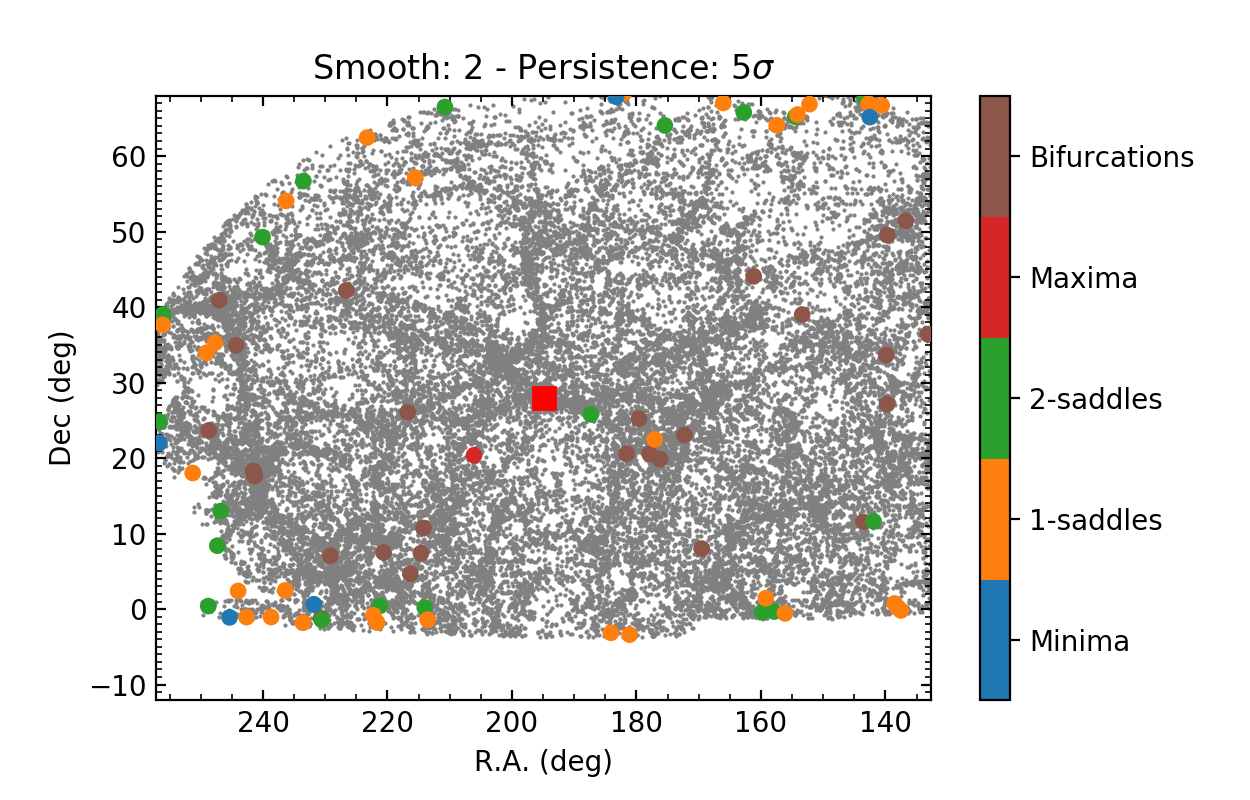}
\caption{Same as Figure \ref{cp_in_coma_slice_smooth_1}. 2-smooth of the density field.}
\label{cp_in_coma_slice_smooth_2}
\end{figure*}

\begin{figure*}
\includegraphics[scale = 0.95]{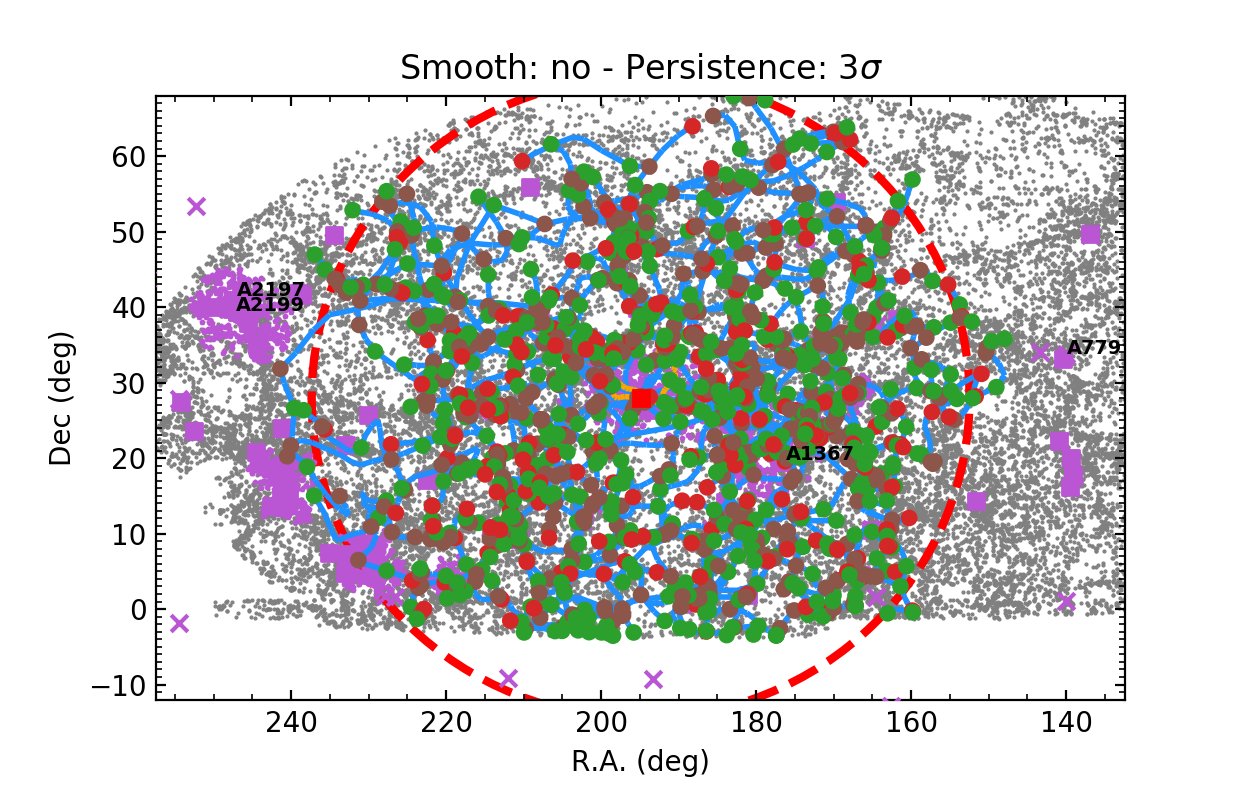}
\includegraphics[scale = 0.95]{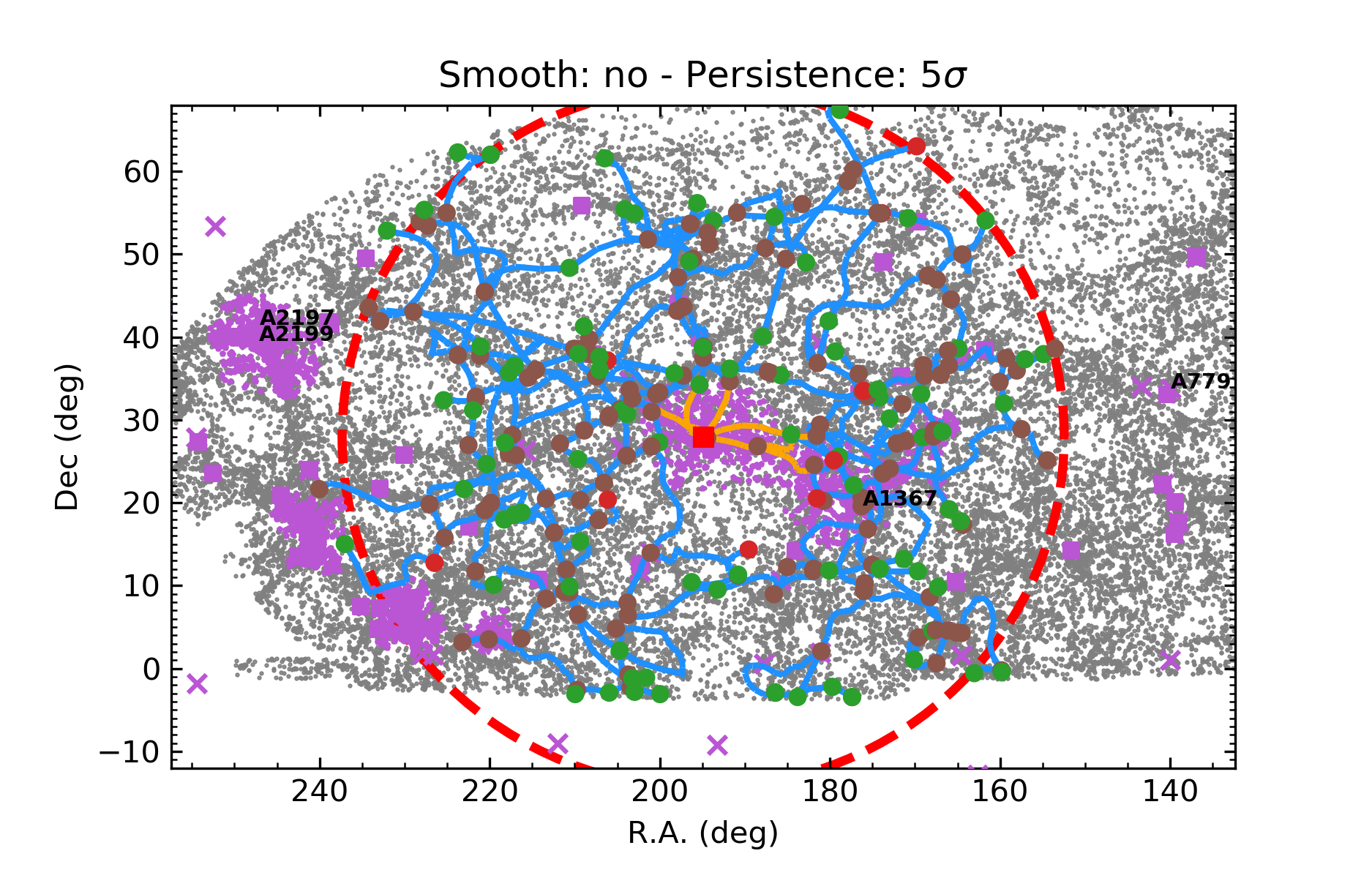}
\caption{Same as Figure \ref{fil_in_coma_sphere_smooth_1}, but for the 1-smooth case.}
\label{fil_in_coma_sphere_no_smooth}
\end{figure*}

\begin{figure*}
\includegraphics[scale = 0.95]{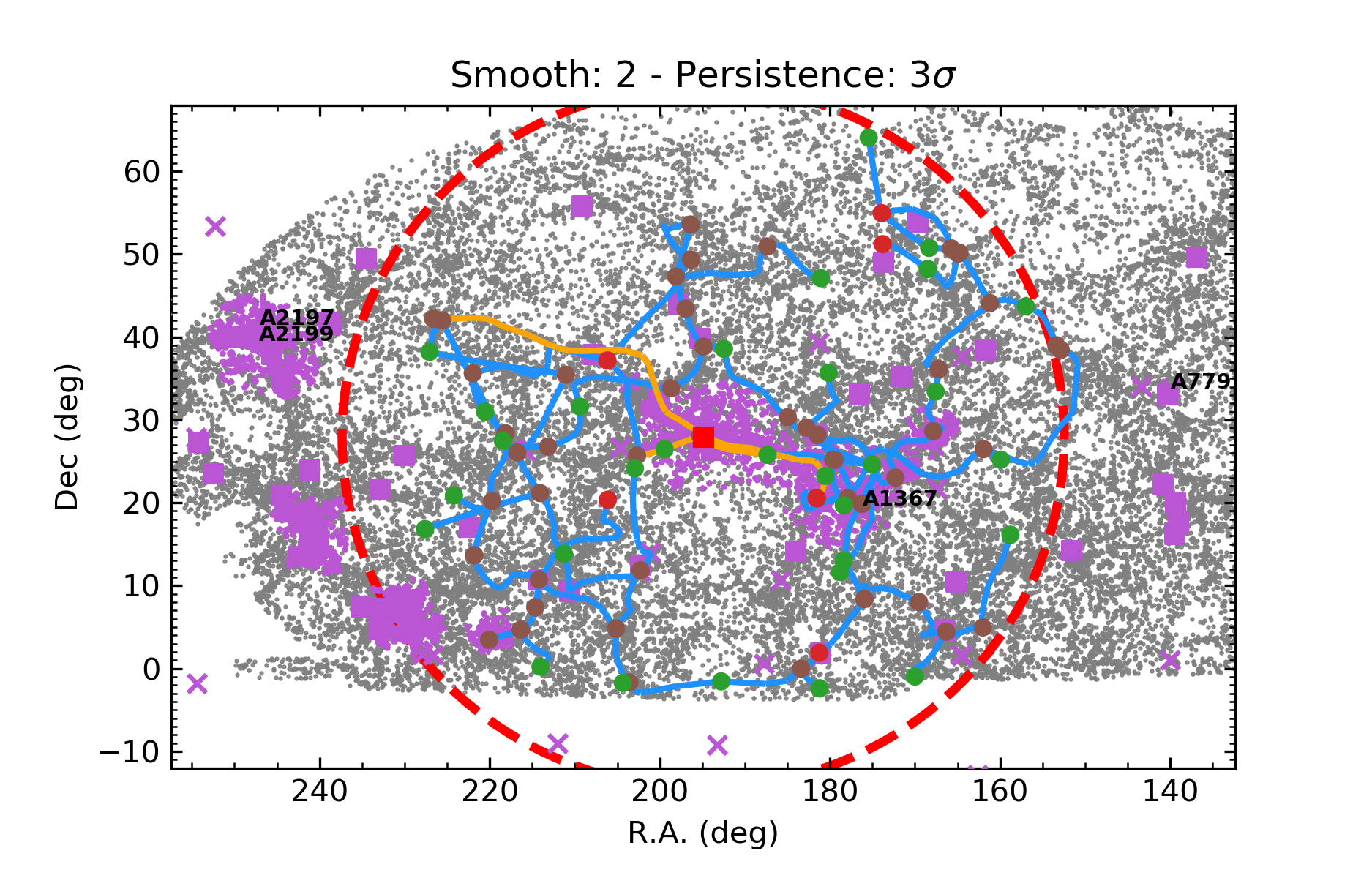}
\includegraphics[scale = 0.95]{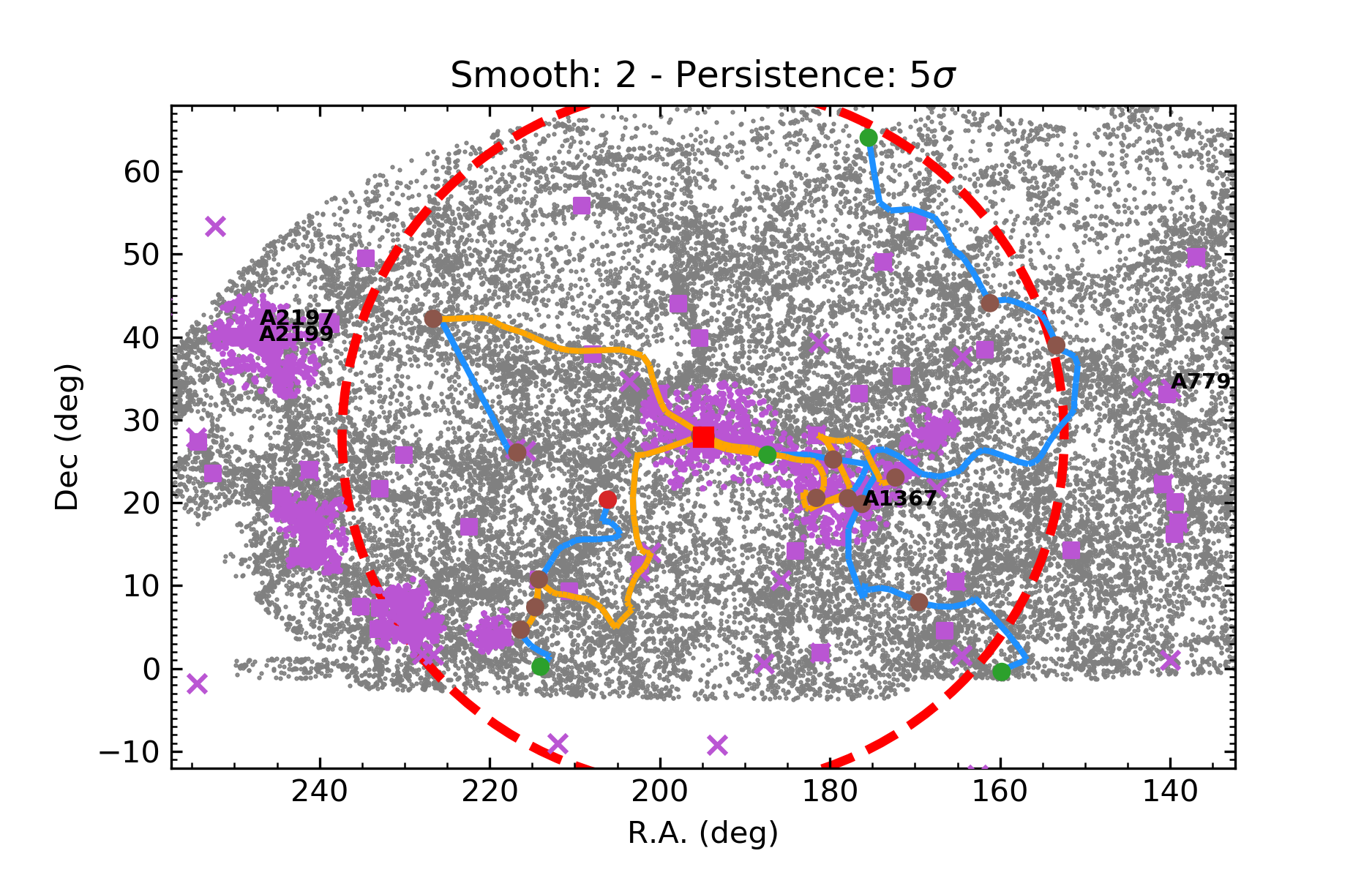}
\caption{Same as Figure \ref{fil_in_coma_sphere_smooth_1}, but for the 2-smooth case.}
\label{fil_in_coma_sphere_smooth_2}
\end{figure*}

\section{Examples of the filament analysis in the SZ}
\label{prof_and_maps}
As stated in Section \ref{results_gasphase}, the input to the RadFil code consists in $20 \times 20 \deg$ patches extracted from the \emph{Planck} $y$-map, which follow the filaments. The RadFil code then extracts profiles of pixel intensity on the map, along lines perpendicular to the filament path on the map. In this appendix we report a few examples of the intermediate steps of the process. Each panel of each figure reports in the top part a cut-out of the \emph{Planck} $y$-map, with rectangles marked which correspond to the position of the patches. The filament path is also reported on the map. In the bottom part of each panel, the individual filaments obtained along lines spaced by one pixel perpendicular to the filament axis and along the filament spine are reported.

\begin{figure*}
\begin{minipage}{\linewidth}
\includegraphics[width = 0.33\linewidth]{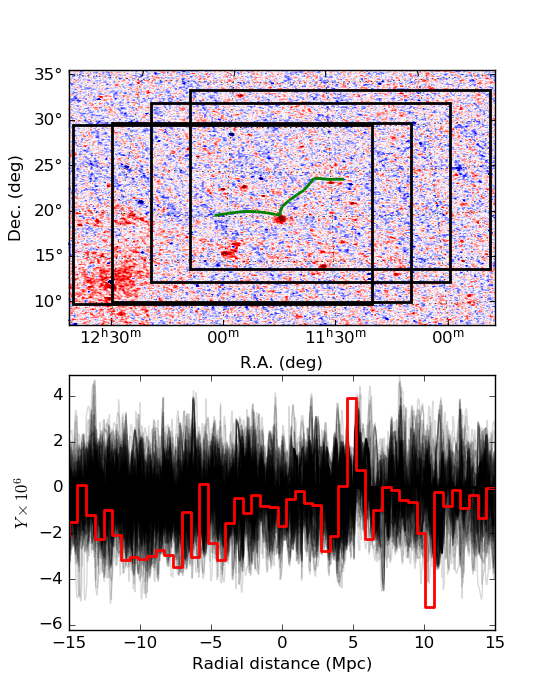}
\includegraphics[width = 0.33\linewidth]{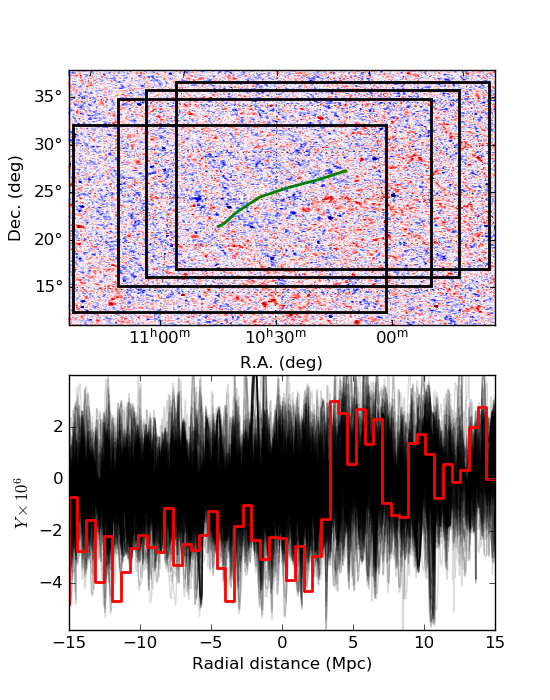}
\includegraphics[width = 0.33\linewidth]{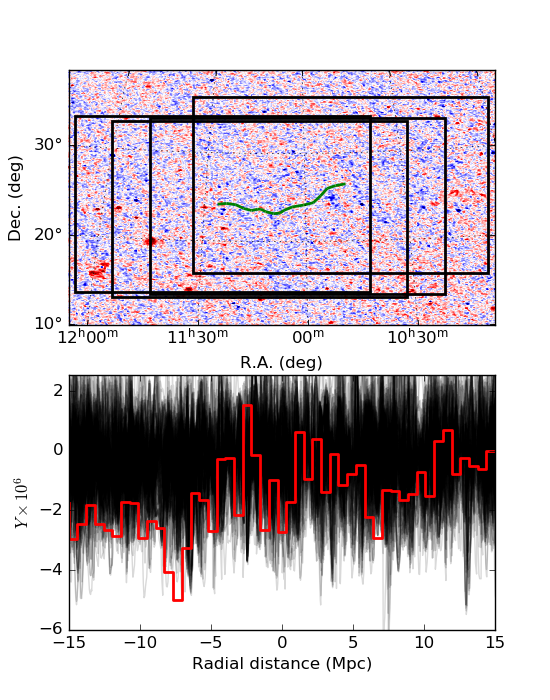}\\
\includegraphics[width = 0.33\linewidth]{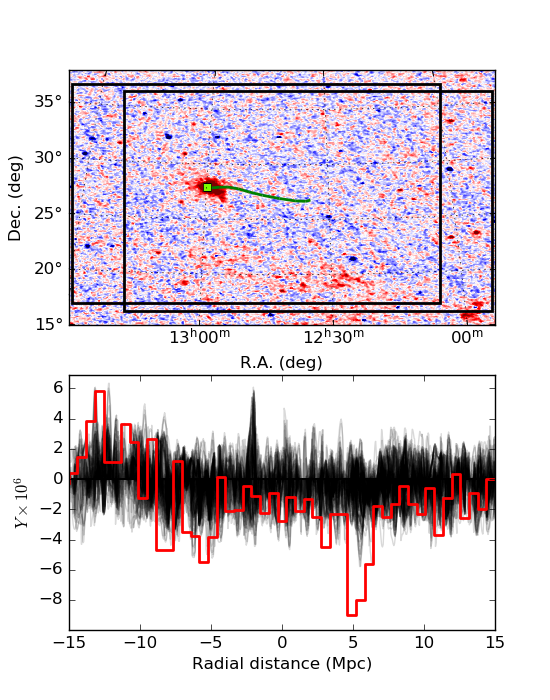}
\includegraphics[width = 0.33\linewidth]{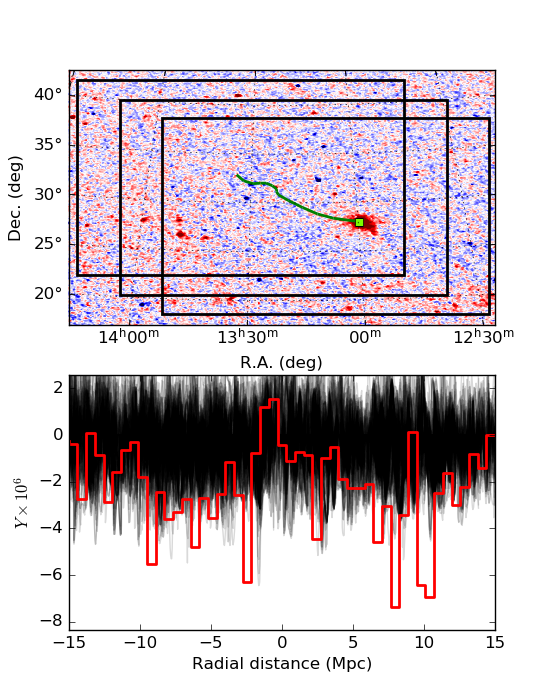}
\includegraphics[width = 0.33\linewidth]{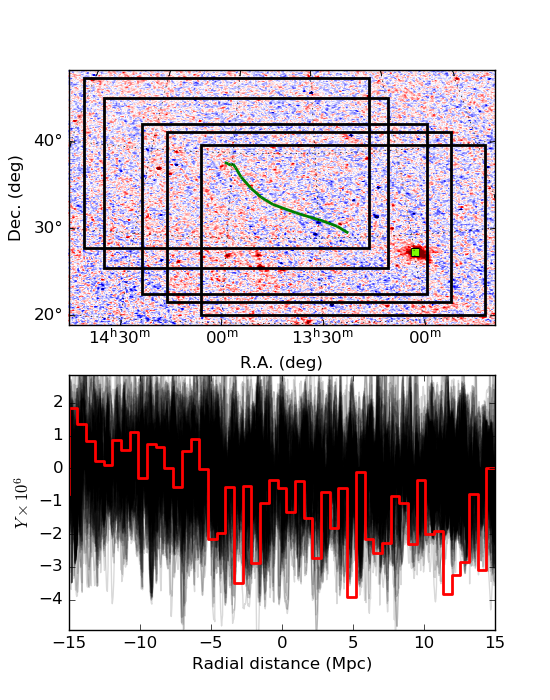}\\
\includegraphics[width = 0.33\linewidth]{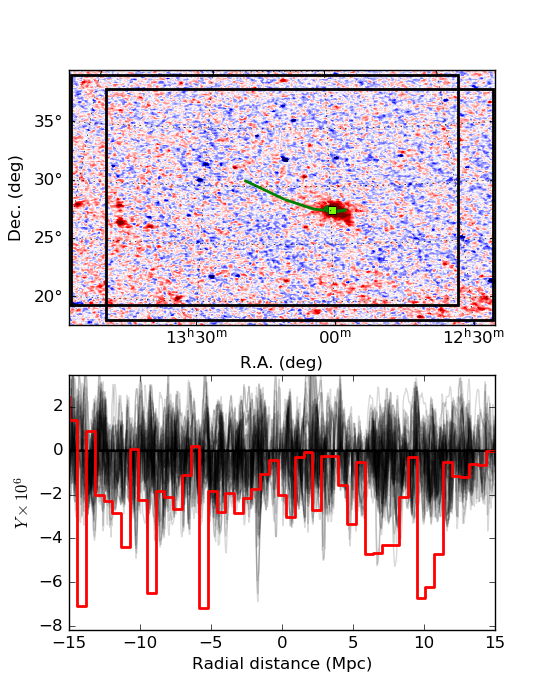}
\includegraphics[width = 0.33\linewidth]{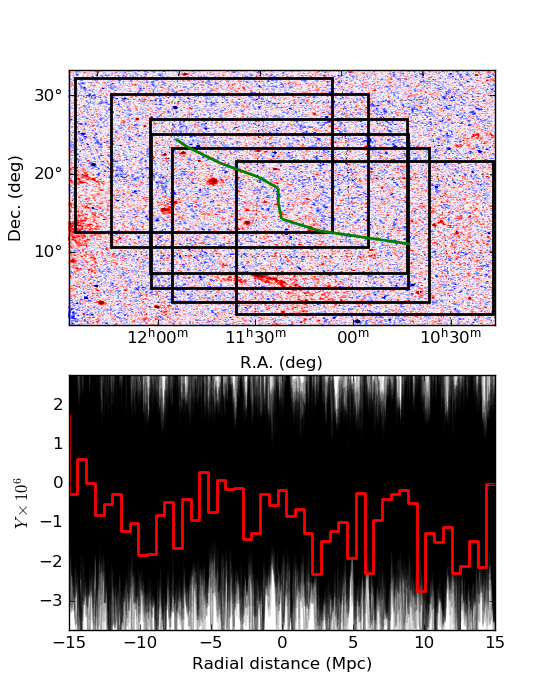}
\includegraphics[width = 0.33\linewidth]{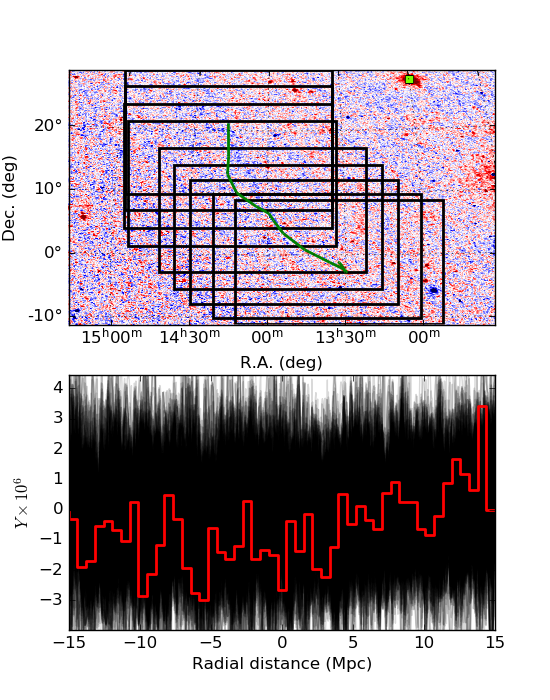}\\
\end{minipage}
\caption{Maps and profiles for a few selected filaments.}
\label{maps_and_profs1}
\end{figure*}

\section{Taking into account the Finger of God effect}
\label{fog}

The Finger of God (FoG) effect is a distortion of the cluster shape in redshift space, occurring when peculiar velocities from the individual galaxies contaminate the redshift measurement \citep[see e.g.][]{Kaiser1987}. Clusters appear as elongated structures in the Line-of-Sight direction. This distortion transforms clusters from almost spherical objects to features that can be easily mistaken for a filament. When applying an algorithm as \disperse to reconstruct the Cosmic Web, the risk is for it to mistake clusters as filaments aligned in the radial direction, therefore biasing the subsequent analysis. As this bias is inevitable unless some operation to mitigate the FoG effect is performed on the data prior to running \disperse, we decided to try to estimate the impact of this problem on our analysis. To this aim, we extracted a slice from the SDSS of $\pm 75 \mathrm{Mpc}$ in the LoS direction centred at the position of the Coma cluster and we run \disperse in 2D in the slice by assuming all galaxies to be at the redshift of Coma.

\begin{figure*}
\centering
\includegraphics[scale = 0.9]{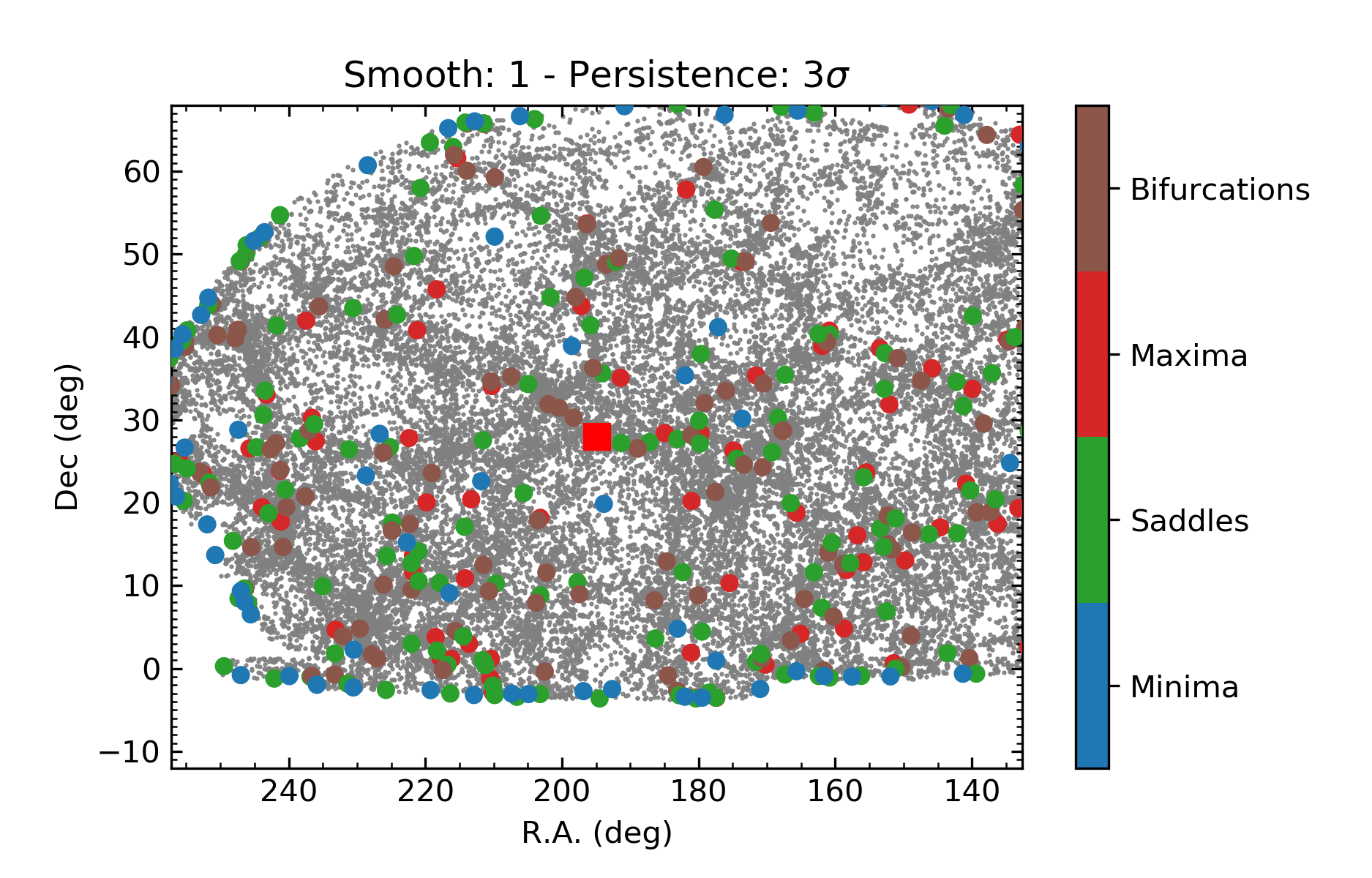}
\caption{Same as Figure \ref{cp_in_coma_slice_smooth_1}, but with \disperse run in 2D.}
\label{crit_disperse_2d}
\end{figure*}

\begin{figure*}
\includegraphics[scale = 0.95]{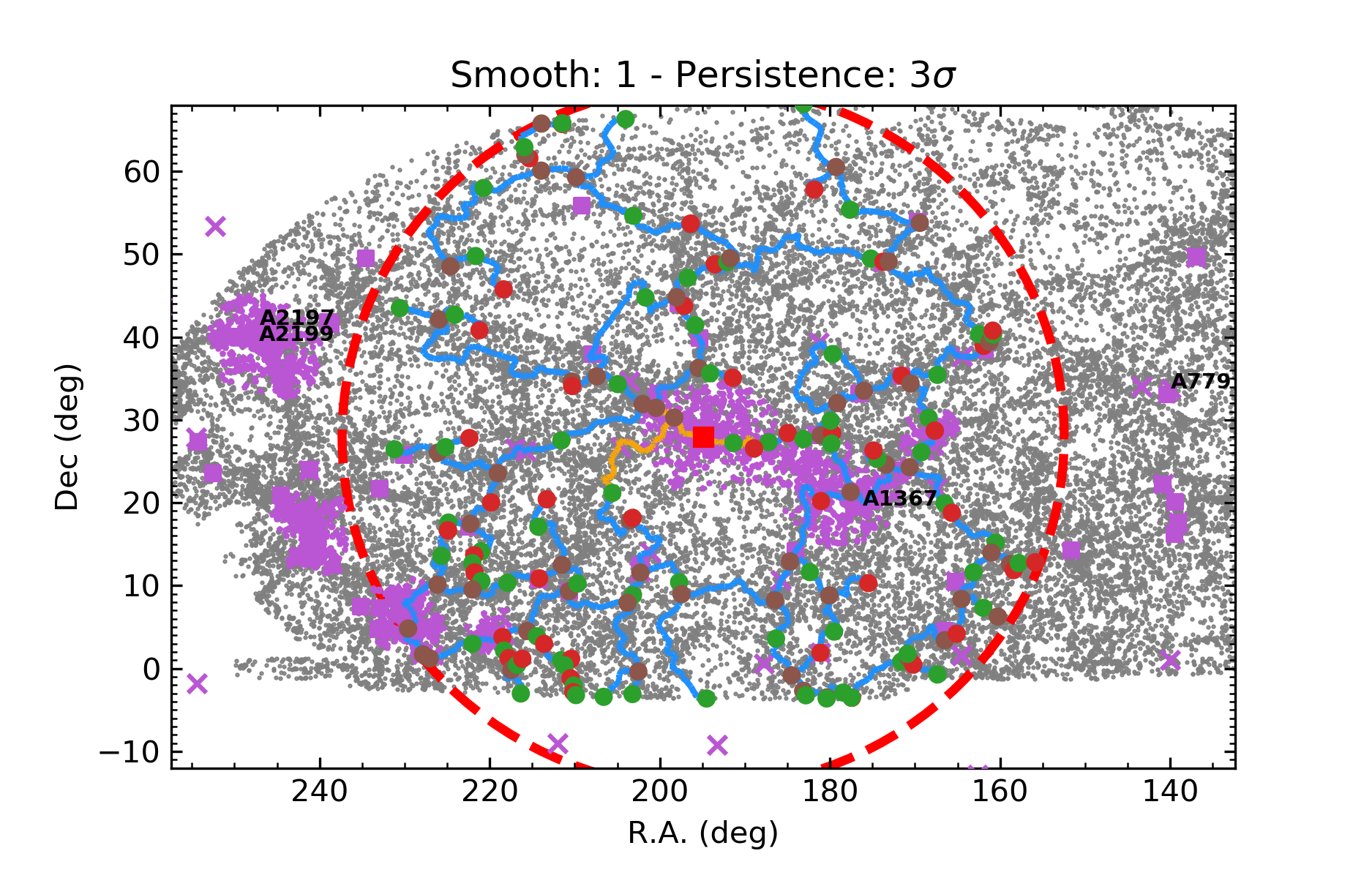}
\caption{Same as Figure \ref{fil_in_coma_sphere_smooth_1}, but with \disperse run in 2D.}
\label{fil_disperse_2d}
\end{figure*}

\begin{figure}
\includegraphics[width=\linewidth]{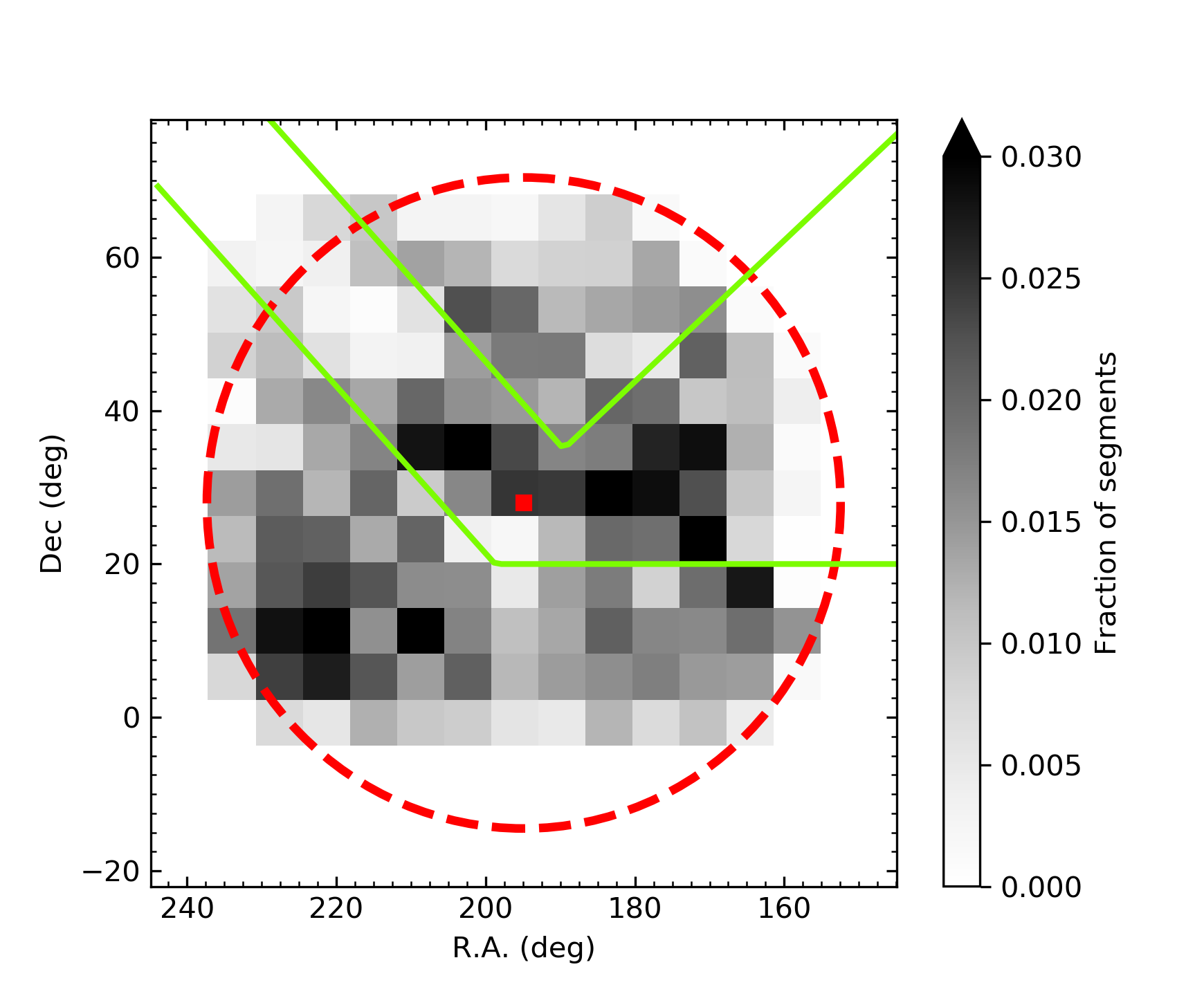}
\caption{Same as Figure \ref{filbeams}, but with \disperse run in 2D.}
\label{filbeams_2d}
\end{figure}

Figures \ref{crit_disperse_2d} and \ref{fil_disperse_2d} show the critical points and the filaments connected to Coma in a 75 Mpc radius region as detected by \disperse in 2D in the SDSS slice. We see how many filaments are actually found in a different position, while some filaments detected in 3D and projected on the plane of the sky have no correspondence in the 2D analysis and new filaments appear in 2D which are not present in 3D. Still, on a statistical basis, the two images do not appear to be too much different. This is somewhat confirmed by the fact that we recover the filament beams in the same position in 2D as in 3D. This result is shown in Figure \ref{filbeams_2d}.

\begin{acknowledgements}
The authors thank Natalya Lyskova and Florian Sarron for sharing data and providing useful insight during the preparation of this manuscript, as well as Andrea Biviano, Clotilde Laigle, and Miguel A. Arag{\'o}n-Calvo for fruitful discussions.

This research has been supported by the funding for the ByoPiC project from the European Research Council (ERC) under the European Union's Horizon 2020 research and innovation programme grant agreement ERC-2015-AdG 695561.

This work makes use of observations obtained with \emph{Planck} (\url{http://www.esa.int/Planck}), an ESA science mission with instruments and contributions directly funded by ESA Member States, NASA, and Canada.

It also made use of the SZ-Cluster Database (\url{http://szcluster-db.ias.u-psud.fr/sitools/clientuser/SZCLUSTER_DATABASE/project-index.html}) operated by the Integrated Data and Operation Centre (IDOC) at the Institut d'Astrophysique Spatiale (IAS) under contract with CNES and CNRS.

Funding for the SDSS and SDSS-II has been provided by the Alfred P. Sloan Foundation, the Participating Institutions, the National Science Foundation, the U.S. Department of Energy, the National Aeronautics and Space Administration, the Japanese Monbukagakusho, the Max Planck Society, and the Higher Education Funding Council for England. The SDSS Web Site is \url{http://www.sdss.org/}.

The SDSS is managed by the Astrophysical Research Consortium for the Participating Institutions. The Participating Institutions are the American Museum of Natural History, Astrophysical Institute Potsdam, University of Basel, University of Cambridge, Case Western Reserve University, University of Chicago, Drexel University, Fermilab, the Institute for Advanced Study, the Japan Participation Group, Johns Hopkins University, the Joint Institute for Nuclear Astrophysics, the Kavli Institute for Particle Astrophysics and Cosmology, the Korean Scientist Group, the Chinese Academy of Sciences (LAMOST), Los Alamos National Laboratory, the Max-Planck-Institute for Astronomy (MPIA), the Max-Planck-Institute for Astrophysics (MPA), New Mexico State University, Ohio State University, University of Pittsburgh, University of Portsmouth, Princeton University, the United States Naval Observatory, and the University of Washington.
\end{acknowledgements}

\bibliographystyle{aa}
\bibliography{coma}

\end{document}